\documentclass[a4paper, 11pt, oneside]{article}
\pdfoutput=1
\usepackage{jheppub}
\usepackage{amssymb}
\usepackage[utf8]{inputenc}
\usepackage{epsfig}
\usepackage{soul}
\usepackage{graphicx}% Include figure files
\usepackage{subfigure}
\usepackage{dcolumn}% Align table columns on decimal point
\usepackage{bm}% bold math
\usepackage[usenames ,dvipsnames]{xcolor}
\usepackage{slashed}

\begin{document}

\title{Strong Dark Matter Self-Interaction \\ from a Stable Scalar Mediator}
\preprint{TUM-HEP/1230/19}
\author[a,b]{Mateusz Duch}
\emailAdd{mateusz.duch@fuw.edu.pl}
\affiliation[a]{Faculty of Physics, University of Warsaw, Pasteura 5, 02-093 Warsaw, Poland}
\affiliation[b]{Physik-Department, Technische Universität München, James-Franck-Straße, 85748 Garching, Germany}
\affiliation[c]{Departamento de Física da Universidade de Aveiro and CIDMA, 
Campus de Santiago, 3810-183 Aveiro, Portugal}
\author[a]{Bohdan Grzadkowski} 
\emailAdd{bohdan.grzadkowski@fuw.edu.pl}
\author[a,c]{Da~Huang}
\emailAdd{da.huang@fuw.edu.pl}
\date{\today}
\abstract{
In face of the small-scale structure problems of the collisionless cold dark matter (DM) paradigm, a popular remedy is to introduce a strong DM self-interaction which can be generated nonperturbatively by a MeV-scale light mediator. However, if such a mediator is unstable and decays into SM particles, the model is severely  constrained by the DM direct and indirect detection experiments. In the present paper, we study a model of a self-interacting fermionic DM, endowed with a light stable scalar mediator. In this model, the DM relic abundance is dominated by the fermionic DM particle which is generated mainly via the freeze-out of its annihilations to the stable mediator. Since this channel is invisible, the DM indirect detection constraints should be greatly relaxed. Furthermore, the direct detection signals are suppressed to an unobservable level since fermionic DM scatterings with a nucleon appear at one-loop level. By further studying the bounds from the CMB, supernovae and BBN on the visible channels involving the dark sector, we show that there is a large parameter space which can generate appropriate DM self-interactions at dwarf galaxy scales, while remaining compatible with other experimental constraints. 
}

%\pacs{95.35.+d, 13.85.Tp, 14.80.-j, 98.70.Sa,}
%\keywords{Dark Matter, Cosmic Rays, AMS-02 Experiment}
\maketitle

%%%%%%%%%%%%%%%%%%%%%%%%%%%%%%%%%%%%%%%%%%%%%%%%%%%%%%%%%%%%%%%%%%%%%%%%%%%%%%%%%%
\section{Introduction}
\label{s1}
Despite great theoretical and experimental efforts in searching for dark matter (DM) during the last several decades~\cite{PDG,Bergstrom:2012fi,Bertone:2004pz,Feng:2010gw}, we still know little about the nature of DM. Until very recently, the most popular DM candidate has been the collisionless cold DM. However, by comparing the $N$-body simulations of such DM with astrophysical observations, a number of discrepancies have been found at the scale of dwarf galaxies, such as the cusp-vs-core problem~\cite{Moore:1994yx,Flores:1994gz,Oh:2010ea,Walker:2011zu} and the too-big-to-fail problem~\cite{BoylanKolchin:2011de,BoylanKolchin:2011dk}. One intriguing possibility to solve these small-scale structure problems is to introduce DM with sufficiently large self-interaction (SIDM)~\cite{deLaix:1995vi,Spergel:1999mh,Vogelsberger:2012ku,Zavala:2012us,Rocha:2012jg,Peter:2012jh, Kaplinghat:2015aga,Tulin:2017ara}, even though there are also some other more conventional solutions~\cite{Governato:2012fa,Brooks:2012vi}. Concretely, the desired DM self-interaction should satisfy $0.1\,{\rm cm^2/g} < \sigma_T/m_{\rm DM} < 10\,{\rm cm^2/g}$, where $\sigma_T$ denotes the so-called momentum transfer cross section and $m_{\rm DM}$ is the mass of the DM particle. On the other hand, DM self-interaction is severely constrained by observations at the galaxy cluster scale to be in the range $\sigma_T/m_{\rm DM} < 1\,{\rm cm^2/g}$~\cite{Clowe:2003tk,Markevitch:2003at,Randall:2007ph,Harvey:2015hha,Kahlhoefer:2013dca, Wittman:2017gxn}. Note that one important difference between DM particles in these two systems are  their average velocities~$v$, with $v= 30$~km/s in dwarfs and $v=1000~{\rm km/s}$ in clusters. Therefore, the observations favor a velocity-dependent DM self-interaction cross section~\cite{Kaplinghat:2015aga}.  

In order to generate so large DM self-scattering cross section, one interesting scenario is to introduce a light ${\cal O}({\rm MeV})$-scale scalar or vector particle to mediate this interaction so that the corresponding DM cross section can be boosted nonperturbatively~\cite{Ackerman:mha,Kaplinghat:2015aga,Feng:2009mn,Buckley:2009in,Loeb:2010gj,Feng:2009hw,Tulin:2012wi,Tulin:2013teo,Cyr-Racine:2015ihg,Aarssen:2012fx,Nozzoli:2016coi}. Note that in this scenario the DM self-interaction cross section increases as the DM velocity becomes small, which is helpful to evade the aforementioned cluster-scale constraint. Furthermore, the observed DM relic density can be naturally obtained by freeze-out of annihilations of DM particles into light mediators~\cite{Feng:2008mu,Foot:2014uba,Berlin:2016gtr,Evans:2017kti}. However, was the mediator unstable and decaying into SM particles, this scenario would face constraints of direct and indirect DM searches. Simplest realizations of the above scenario are models with weak-scale fermionic DM and MeV-scale vector or scalar mediator. For the case with a vector mediator, it was shown in Ref.~\cite{Bringmann:2016din,Cirelli:2016rnw} that the model was ruled out by indirect detection constraints from Cosmic Microwave Background (CMB)~\cite{Ade:2015xua,Padmanabhan:2005es,Slatyer:2015jla, Slatyer:2015kla,Poulin:2016anj}, Fermi-LAT~\cite{Ackermann:2015zua} and AMS-02~\cite{AMS1, AMS2, Bergstrom:2013jra,Hooper:2012gq,Ibarra:2013zia}. On the other hand, in the model with a scalar mediator, even though the DM indirect detection constraints can be avoided since the fermionic DM annihilation is $p$-wave dominated~\cite{Kahlhoefer:2017umn}, the DM direct detection upper bounds~\cite{Aprile:2018dbl} imply the longevity of the scalar mediator, which would modify the primordial abundances of light elements during Big-Bang nucleosynthesis (BBN)~\cite{Kainulainen:2015sva,Kaplinghat:2013yxa,Hufnagel:2018bjp}. Though, there have already been many possible solutions proposed to avoid such tensions for models with scalar and vector mediators~\cite{Duch:2017khv,Duch:2017qjc,Bernal:2015ova,Blennow:2016gde,Baldes:2017gzu, Kahlhoefer:2017umn,Bringmann:2018jpr,Ma:2017ucp,Barman:2018pez,Caputo:2019wsd,Ahmed:2017dbb,Duerr:2018mbd}.

In the present paper, inspired by Refs.~\cite{Ma:2017ucp,Ahmed:2017dbb,Duerr:2018mbd}, we construct and study a DM model in which the dominant DM component is a fermionic particle with a \textit{stable} light scalar mediator. The stable mediator constitutes a subdominant DM by the freeze-out of its annihilation into additional light scalar particles. In this model, the observed DM relic density is mainly obtained by the freeze-out of the annihilation of a fermionic DM pair into the scalar mediators, which cannot be probed by DM indirect detection. Moreover, the fermionic DM is also free from any DM direct search constraints, since its nuclear recoils appear at one-loop level. Thus, the model is expected to be less constrained compared to the counterpart with an unstable scalar mediator, and has the potential to reconcile the aforementioned conflict among different experiments. However, as will be shown below, this model might be experimentally tested and constrained by observations of BBN~\cite{Scherrer:1987rr,Hufnagel:2017dgo} and CMB~\cite{Padmanabhan:2005es,Slatyer:2015jla, Slatyer:2015kla,Poulin:2016anj}, since the processes involving dark-sector particles can still leave their tracks in modifications of the primordial abundances of light elements and it can change the CMB power spectrum. Therefore, the main question in the following is whether we can find the parameter space which can produce the desired strong DM self-interactions, while still be consistent with the current experimental bounds.

The paper is organized as follows. In Sec.~\ref{Sec_model}, we briefly introduce our model and clarify notation and conventions. Sec.~\ref{Sec_RelDens} is devoted to the calculation of the DM relic density. The constraints from CMB, supernovae and BBN on the dark Higgs properties are discussed in Sec.~\ref{Sec_h2}.  In Sec.~\ref{Sec_DD}, we show the analytic expressions of nuclear recoil cross sections for both DM components in the limit of zero momentum transfer, showing that they are effectively invisible under current experimental status. In Sec.~\ref{Sec_DMID}, we discuss DM indirect detection constraints for visible channels involving dark sector particles. Discussion of the calculation of DM self-interactions is given in Sec.~\ref{Sec_DMSI}. Then we show our numerical studies in Sec.~\ref{Sec_NumRes}. Finally, we present our conclusions in Sec.~\ref{Sec_Conc}.

%%%%%%%%%%%%%%%%%%%%%%%%%%%%%%%%%%%%%%%%%%%%%%%%%%%%%%%%%%%%%%%%%%%%%%%%%%%%%%%%%%
\section{The Model}
\label{Sec_model}
Our model is a simple extension of the SM by including a Dirac fermion $\chi$ and two real scalars $S$ and $\phi$. We firstly impose the following stabilizing $Z_4$ symmetry on the model~\cite{Cai:2015zza}:
\begin{eqnarray}
\chi \to i\chi\,, \quad\quad \quad S \to -S\,,
\end{eqnarray}
with other fields being neutral. {The symmetry implies that the SM scalar potential $V_H=\frac{1}{2} \mu_H^2 |H|^2 + \frac{1}{4}\lambda_H |H|^4$ is extended to contain the terms with $S$}
\begin{eqnarray}
V_{S} =  \frac{1}{2} \mu_S^2 S^2 + \frac{1}{4}\lambda_S S^4 + \frac{1}{2}\kappa_{HS} S^2 |H|^2
\end{eqnarray}
and the following terms involving $\chi$
\begin{eqnarray}\label{YchiS}
V_{\chi S} = m_\chi \bar{\chi}\chi + \frac{g_Y}{2} S (\bar{\chi^c}\chi+\bar{\chi}\chi^c)\,.
\end{eqnarray}
%After the spontaneous breaking of the electroweak (EW) gauge symmetries, the SM Higgs obtains a vacuum expectation value $\langle H \rangle = (0,v_H/\sqrt{2})^T$, and the physical mass of $S$ is $m_S^2 = \mu_S^2 + \kappa v_H^2/2$.
Since we assume $\langle S \rangle =0$, the stabilizing symmetry remains unbroken. Note also that the above Lagrangian is invariant under its subgroup of this symmetry: $\chi \to -\chi$~\cite{Cai:2015zza}. Therefore, when mass of the scalar $S$ is below half of the Dirac fermion $\chi$ mass, both particles are stable and can contribute to the final DM relic density. Since our intention is to enhance the fermionic DM self-interactions by an exchange of the scalar $S$, we will consider the case with $\chi$ much heavier than $S$. Concretely, $\chi$ is assumed to be at the electroweak scale with its mass $m_\chi \sim {\cal O}(1~{\rm GeV} \div 100~{\rm GeV})$, while the mass of $S$ will be of ${\cal O}({\rm MeV})$. However, for such a small mass of $S$, the annihilation of $S$ into SM particles are typically inefficient to reduce its relic abundance to be subdominant. Thus, we need to introduce an additional real scalar $\phi$ so that there is an extra annihilation channel for $S$ to deplete its abundance. In order to avoid unnecessary parameters in the Lagrangian, we impose an extra $Z_2$ symmetry: $\phi \to -\phi$, which implies the following potential terms
\begin{eqnarray}
V_\phi = -\frac{\mu^2_\phi}{2}\phi^2 + \frac{1}{4}\lambda_\phi \phi^4 +\frac{1}{4} \kappa_{S\phi} S^2 \phi^2 + \frac{1}{2}\kappa_{H\phi} |H|^2 \phi^2\,,
\end{eqnarray}
Note that the latter $Z_2$ symmetry would be broken by the VEV of $\langle\phi\rangle = v_\phi$ so that the perturbation $\varphi = \phi-v_\phi$ can  mix with the neutral component $h$ of the Higgs doublet which is defined as $H \equiv (0, (v_H+h)/\sqrt{2})^T$ in the unitary gauge with $v_H = 246$~GeV. 
% As a result, $\varphi$ is unstable and can decay into lighter SM particles.
% \bg{Below we use $\lambda_H$, so the SM part of the potential should also be explicitly written} 

By minimizing the total scalar potential, we can determine non-zero VEVs of $H$ and $\phi$ solving the following two equations:
\begin{eqnarray}
-\mu^2_H + \lambda_H v_H^2 + \frac{1}{2} \kappa_{H\phi} v_\phi^2 =0\,,\nonumber\\
-\mu_\phi^2 + \lambda_\phi v_\phi^2 + \frac{1}{2} \kappa_{H\phi}v_H^2 = 0 \,.
\end{eqnarray} 
We further expand the potential written in terms of perturbation fields $S$, $h$, and $\varphi$ up to the second order, determining the mass squared of $S$ as $m_S^2 = \mu_S^2 + (\kappa_{HS} v_H^2 + \kappa_{S\phi} v_\phi^2)/2$ and the following mass squared matrix for $h$ and $\varphi$:
\begin{eqnarray}
{\cal M}^2_{h\varphi} = \left(\begin{array}{cc}
2\lambda_H v_H^2 & \kappa_{H\phi} v_H v_\phi \\
\kappa_{H\phi} v_H v_\phi & 2\lambda_\phi v_\phi^2 
\end{array}\right)\,.
\end{eqnarray}
We can diagonalize the matrix above defining the mass eigenstate $h_{1,2}$ in terms of the perturbations and mixing angle $\theta$ as follows
\begin{eqnarray}
\left(\begin{array}{c}
h \\
\varphi
\end{array}\right) = \left(\begin{array}{cc}
c_\theta & -s_\theta \\
s_\theta & c_\theta
\end{array}\right) \left(\begin{array}{c}
h_1 \\
h_2
\end{array}\right)\ ,
\end{eqnarray}
where $s_\theta \equiv \sin\theta$ and $c_\theta \equiv \cos\theta$. Denoting the masses of $h_{1,2}$ by $m_{1,2}$, the following relations between parameters hold
\begin{eqnarray}
\lambda_H = \frac{c_\theta^2 m_1^2 + s_\theta^2 m_2^2}{2v_H^2}\,, \quad \lambda_\phi = \frac{s_\theta^2 m_1^2 + c_\theta^2 m_2^2}{2v_\phi^2}\,,\quad \kappa_{H\phi} = \frac{(m_1^2-m_2^2)s_\theta c_\theta}{v_H v_\phi}\,.
\end{eqnarray}
If we assume that $h_1$ is the SM-like Higgs state, then $m_1 = 125$~GeV and the model can be parametrized by the following 9 free parameters
\begin{eqnarray}
m_2\,, m_S\,, m_\chi\,, s_\theta\,, \lambda_\phi\,, \kappa_{HS}\,,\kappa_{H\phi}\,, \kappa_{S\phi}\,, g_Y\,.
\end{eqnarray}

%%%%%%%%%%%%%%%%%%%%%%%%%%%%%%%%%%%%%%%%%%%%%%%%%%%%%%%%%%%%%%%%%%%%%%%%%%%%%%%%%%%%%%%%%
\section{Dark Matter Relic Density}
\label{Sec_RelDens}
In the present model, there are two stable particles: the heavy DM candidate $\chi$ and the stable mediator $S$, both of which will contribute to the final DM relic abundance. As mentioned before, the stable mediator $S$ should be hierarchically lighter than the fermionic DM $\chi$, {\it i.e.}, $m_S \ll m_\chi$, so that the self-interactions of $\chi$ can be non-perturbatively enhanced to the level large enough to solve the small-scale structure problems. In this section, we would like to discuss how the DM relic density can be obtained via the freeze-out mechanism in this scenario.

The number densities of the two DM components, $n_\chi$ and $n_S$, can be yielded by solving the following two coupled Boltzmann equations:
\begin{eqnarray}\label{Boltzmann}
\frac{d n_\chi}{dt} + 3 H n_\chi &=& -(\frac{1}{2}\langle \sigma v \rangle_{\chi\bar{\chi} \to SS} + \frac{1}{4} \langle \sigma_{\rm SA} v \rangle_{\chi{\chi} \to S h_{2}} + \frac{1}{4} \langle \sigma_{\rm SA} v\rangle_{\bar{\chi}\bar{\chi} \to S h_{2}}) (n_\chi^2 - n_\chi^{{\rm eq}\,2}) \nonumber\\
\frac{d n_S}{dt} + 3H n_S &=& -\langle \sigma v \rangle_{SS \to h_2 h_2} (n_S^2 - n_S^{{\rm eq}\,2}) \nonumber\\
&& + \Big(\langle \sigma v \rangle_{\chi\bar{\chi} \to SS} + \frac{1}{4}\left[ \langle \sigma_{\rm SA} v \rangle_{\chi\chi \to S h_{2}} + \langle \sigma_{\rm SA} v \rangle_{\bar{\chi} \bar{\chi} \to S h_{2}} \right]  \Big) (n_\chi^2 - n_{\chi}^{{\rm eq}\,2})\nonumber\\
&& -\langle \sigma v \rangle_{\chi S \to \bar{\chi} h_{2} (\bar{\chi} S\to \chi h_{2})} (n_S - n_S^{\rm eq})n_{\chi}
\end{eqnarray}
where $n_{\chi}$ is the total number density of both $\chi$ and $\bar{\chi}$, while $n_S$ that of the light scalar mediator $S$. On the right-hand side of both equations, we have only shown the most important processes for the determination of the DM relic densities, including the dominant annihilations for both components ($\chi \bar{\chi} \to SS$ for $\chi$ and $SS \to h_2 h_2$ for $S$), the semi-annihilation (SA) process $\chi\chi(\bar{\chi}\bar{\chi}) \to Sh_{2}$, and the conversion process $\chi S \to \bar{\chi} h_{2}$ ($\bar{\chi} S \to \chi h_{2})$~\cite{Cai:2015zza}.
\begin{figure}[]
\centering
\includegraphics[width = \linewidth]{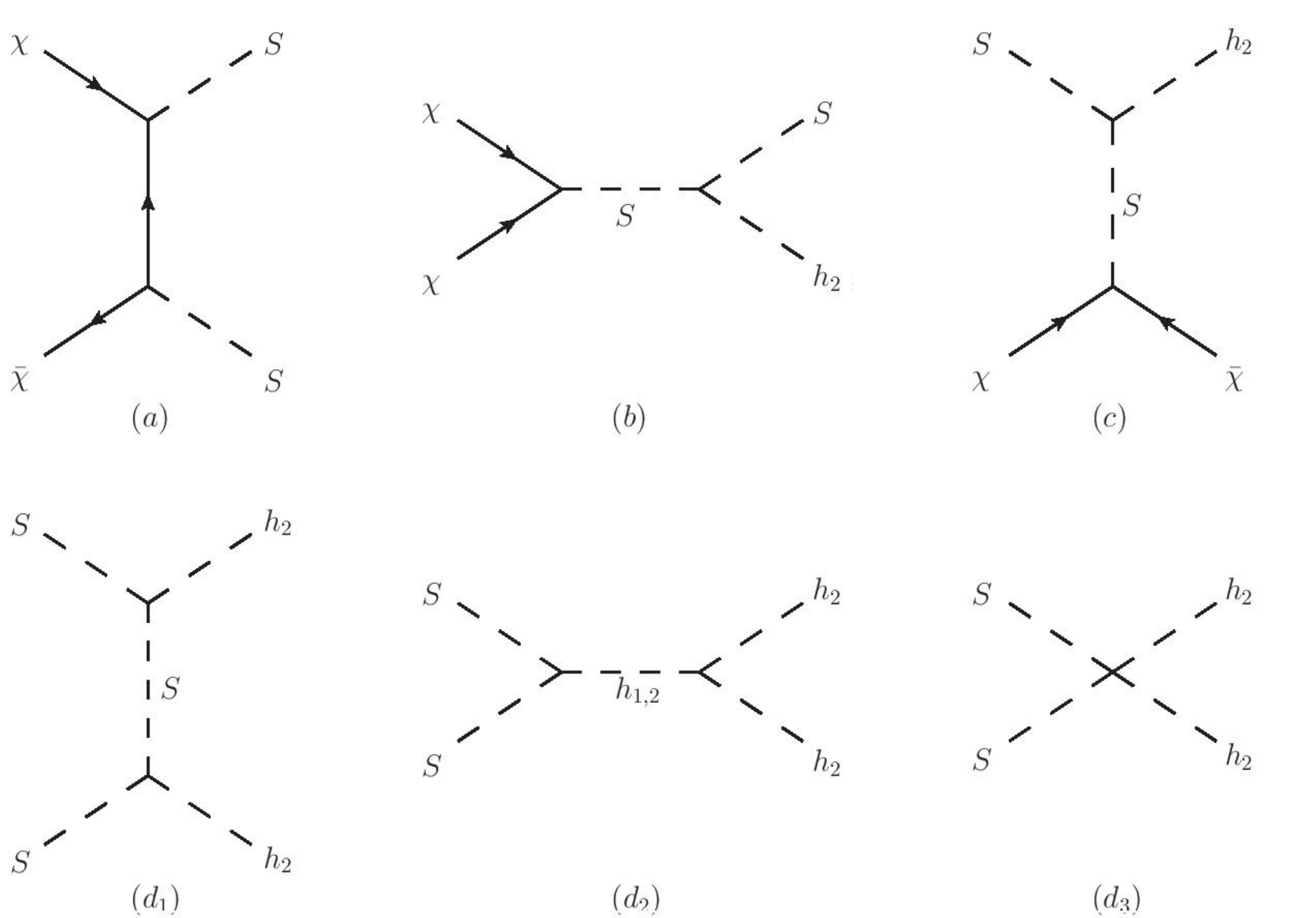}
\caption{Feynman diagrams for the processes involved in the calculation of DM relic density: (a) $\chi\bar{\chi} \to SS$; (b) semi-annihilation $\chi\chi \to S h_{2}$; (c) conversion $\chi S \to \bar{\chi} h_2$; ($d_1$-$d_3$) $SS \to h_2 h_2$. For the semi-annihilation process $\bar{\chi} \bar{\chi} \to S h_{2}$ and the conversion one $\bar{\chi} S \to \chi h_2$, the corresponding Feynman diagrams can be obtained by reversing the directions of fermionic lines in (b) and (c).}\label{dmrd}
\end{figure}
The corresponding Feynman diagrams are listed in Fig.~\ref{dmrd}. Other processes are suppressed either by the small mixing angle $s_\theta$ like $\chi S \to \bar{\chi} h_1$ or by the loop factors such as the $\chi\bar{\chi}$ annihilations into SM particles.

Note that when the freeze-out of the heavy fermionic DM $\chi$ takes place, cross sections for $\chi\bar{\chi}$, $\chi\chi$ and $\bar\chi\bar{\chi}$ annihilations are dominated by the $p$-wave contribution~\cite{Kahlhoefer:2017umn}. By further taking into the multiple exchange effects of the light mediator $S$, we should also multiply them by the following $p$-wave Sommerfeld enhancement factor~\cite{Sommerfeld,Cassel:2009wt,Iengo:2009xf,Slatyer:2009vg, Tulin:2013teo}
\begin{eqnarray}
S_p = \frac{(c-1)^2+4 a^2 c^2}{1+4 a^2 c^2} S_s\,,
\end{eqnarray}
where the factor $S_s$ is the Sommerfeld factor for $s$-wave annihilations with its explicit form given by
\begin{eqnarray}
S_s = \frac{\pi}{a} \frac{\sinh(2\pi a c)}{\cosh(2\pi a c)-\cos(2\pi\sqrt{2\pi\sqrt{c-(ac)^2}})}\,.
\end{eqnarray}
Here, we have defined $a \equiv v/(2\alpha_X)$ and $c\equiv 6\alpha_X m_\chi/(\pi^2 m_S)$, with $\alpha_X \equiv g_Y^2/(4\pi)$ and $v$ is the relative velocity between $\chi$ and $\bar{\chi}$. However, we find that this effect is so small that it can be neglected in the final prediction of relic densities of both DM components.  

Furthermore, as will be discussed in Sec.~\ref{Sec_h2}, the dark sector should decouple from the visible SM sector at least before the QCD phase transition at $T_{\rm QCD} \approx 100 \div 200$~MeV, here after we will adopt $T_{\rm dec}\approx 500$~MeV. After decoupling the two sectors evolve independently with their own temperatures~\cite{Duerr:2018mbd}. Under the assumption of the entropy conservation in each sector, the temperature ratio between the dark and SM sectors can be estimated as a function of the SM plasma temperature $T$
\begin{eqnarray}
\xi (T) \equiv \frac{T_D(T)}{T} = \left(\frac{g_{*S}(T)}{g_{*S}(T_{\rm dec})} \frac{g^D_{*S}(T_{\rm dec})}{g^D_{*S}(T_D)}\right)^{1/3}\,,
\end{eqnarray}   
where $g_{*S}(T)$ and $g_{*S}(T_D)$ are the relativistic degrees of freedom relevant for the entropy in the visible and dark sectors, respectively. 
In the present model with the desired DM masses, the stable scalar mediator $S$ always freezes out after the decoupling, while the relic density of the fermionic DM $\chi$ can be also affected as long as its mass is lighter than 12 GeV~\cite{Duerr:2018mbd}. In order to account for the impact of this thermal decoupling of the dark sector on the DM relic abundances quantitatively, we follow Ref.~\cite{Feng:2008mu} simply multiplying the relic abundances of $S$ and/or $\chi$ obtained assuming equal temperatures in both sectors by the correction factor $\xi(T_{\rm f})$, where $T_{\rm f}$ denotes the freeze-out temperatures of the corresponding stable particles. 

In our work, we numerically solve the coupled Boltzmann equations in Eq.~(\ref{Boltzmann}) using the modified \texttt{MicrOMEGAs v4.3.5} code~\cite{Belanger:2006is,Belanger:2014vza} which takes into account the aforementioned Sommerfeld enhancement and decoupling effects of the hidden sector. As noted in Ref.~\cite{Duerr:2018mbd}, when calculating the final relic abundance of the subdominant component $S$, the $\chi\bar{\chi}$ annihilation into a pair of $S$ and the semi-annihilation process can play crucial roles, even though they already froze out and cannot further modify the abundances of $\chi$ and $\bar{\chi}$.

\section{Constraints on Dark Higgs Boson Decays}
\label{Sec_h2}

Since the dark sector is assumed to be in thermal equilibrium with the SM particles at early times, the dark Higgs boson $h_2$ would exist abundantly in the early Universe. In order to facilitate effective annihilation of the mediator, we are going to consider a light $h_2$ that can only decay into $e^+ e^-$ and $\gamma\gamma$. In the low mass range, the main bounds on its lifetime and mixing angle come from the beam dump experiments, supernovae, BBN and CMB.

In particular, when $m_2 > 2 m_e$, $h_2$ mainly decays into $e^+ e^-$, while for the $h_2$ mass below this threshold {only the $\gamma\gamma$ mode induced at one-loop is accessible. However, if} the respective decays are effective at redshifts $z \lesssim 2\times 10^6$, the produced electromagnetic energy cannot fully thermalize with the background plasma and thus induces the spectral distortion in the CMB~\cite{Zeldovich:1969ff,Hu:1993gc,Chluba:2011hw}. It is shown in Ref.~\cite{Poulin:2016anj} that this constraint can be transformed into the limit on the $h_2$ lifetime $\tau_{h_2} < 10^5$~s while the $\gamma\gamma$ channel always gives the lifetime much larger than this bound.
% \bg{Since there is only one possibility for the decays, the proceeding two phrases are not quite appropriate in its present form.} 

Consequently, we focus on the first region $m_2 > 2 m_e$ where the decay rate of $h_2$ is given by
\begin{eqnarray} 
\Gamma(h_2 \to e^+ e^-) = \frac{s_\theta^2}{8\pi} \frac{m_e^2}{v_H^2} \frac{(m_2^2-4m_e^2)^{3/2}}{m_2^2}\,.
\end{eqnarray}
In the further discussion, we choose the value $m_2 = 1.5$~MeV as in Ref.~\cite{Duerr:2018mbd}. We note that, in contrary to the model presented there, in our scenario the mixing angle $s_\theta$ is a free parameter, therefore $h_2$ lifetime is independent from the Higgs portal coupling $\kappa_{H\phi}$, the coupling of mediator to the DM $g_Y$ or its mass $m_S$.

% The limit on the lifetime implies $s_\theta > 2.55\times 10^{-7}$.
The mixing angle is limited from above by beam dump experiments. The CHARM Collaboration looks for axion-like particles \cite{Bergsma:1985qz} using a $400$~GeV proton beam and a copper target. Their searches result in the bound $\theta^2<10^{-6}$ for a scalar of mass $1.5$~MeV~\cite{Clarke:2013aya,Krnjaic:2015mbs}.

{Another set of constraints arise from the supernova (SN) explosions~\cite{Turner:1987by,Frieman:1987ui,Burrows:1988ah,Ishizuka:1989ts,Essig:2010gu}. The $h_2$ can be radiatively produced off nucleons resulting in the extra energy loss and significant shortening of the duration of the neutrino pulse during the SN explosion. Currently, the main bound is given by observation of the SN1987a. According to Fig.~4 in Ref.~\cite{Krnjaic:2015mbs}, for $h_2$ with mass 1.5~MeV, the SN1987a limits the Higgs mixing angle to be $s_\theta^2 \gtrsim 10^{-9}$ or $s_\theta^2 \lesssim 10^{-12.3}$.} 

Furthermore, the presence of long-lived $h_2$ may spoil the sucessful prediction of BBN by two main effects~\footnote{If we considered heavier dark Higgses, light nuclei could also be photodisintegrated by the $h_2$ decay products with energy above the threshold $E_{\rm dis} = 2.2$~MeV. This effect is relevant only when $h_2$ lifetime exceeds $10^4$~s.}. First, an additional relativistic degrees of freedom (dofs) boost the cosmological expansion rate and alter the ratio of proton and neutron number densities when they leave thermodynamic equilibrium. Secondly, $h_2$ decays produce entropy in the visible sector and lead to a modified time dependence of baryon-to-photon ratio that strongly influences the calculation of light element abundances. Moreover, the production of extra entropy also lowers the neutrino-to-photon temperature ratio with respect to the SM value and results in the reduced value of the effective number of neutrino species $\Delta N^{\rm CMB}_{\rm eff}$ at recombination that is constrained by the Planck data \cite{Aghanim:2018eyx} 

The BBN and CMB bounds on the electromagnetically decaying dark sector particles were considered recently in Ref. \cite{Hufnagel:2018bjp}. As presented in fig. 9 therein, BBN limits the lifetime of $h_2$ to the range $\tau_{h_2}\lesssim 10$~s when dark Higgs has mass $m_{h_2}\gtrsim 1$~MeV and remains in the chemical equilibrium until temperature $T^{\rm cd}\sim 1$~MeV~\footnote{The bounds of fig. 9 in \cite{Hufnagel:2018bjp} were calculated assuming $T^{\rm cd}=10$ GeV. For another decoupling temperature, the limits should be rescaled by a factor $g_s(T^{\rm cd})/g_s(10\;{\rm GeV})$ describing the change in the number of relativistic degrees of freedom.}. However, for $\tau_{h_2}\sim 10$~s, the CMB bounds turn out to be stronger (cf. fig. 4 of \cite{Hufnagel:2018bjp}, see also \cite{Fradette:2017sdd}) and exclude thermally produced dark Higgs, unless it decouples earlier and dilutes with respect to photons that are heated with decreasing number of relativistic degrees of freedom in the visible sector.

Taking into account the supernovae bounds, for $s_\theta^2<10^{-12.3}$, the lifetime of $h_2$ is in the range $\tau_{h_2}>1.3 \times 10^4$~s that is excluded by BBN and CMB. On the other hand, for $s_\theta^2>10^{-9}$ the dark Higgs decays earlier ($\tau_{h_2}<6.5$~s) and the bounds may be avoided  when its density is small enough. 
In the following, we will focus on the marginal value of the mixing angle $s_\theta^2=10^{-9}$ and show that CMB bound are indeed satisfied when the dark sector decouples at $T_{\rm dec}=500$~MeV thus certainly earlier then the onset of QCD phase transition occurring at $T_{\rm QCD} \approx 100 \div 200$~MeV. After the thermal decoupling, the comoving entropies in both the SM sector $s_{\rm SM}$ and the dark sector $s_D$ are conserved. Since then, the chemical decoupling of the stable mediator $S$ and $h_2$ in the hidden sector happens when $S$ is non-relativistic, its contribution to $s_D$ is negligible with respect to that of the lighter $h_2$. It follows that
\begin{eqnarray}
 \left.\frac{n_{h_2}}{n_\gamma}\right|_{T^{\rm cd}} &=& \left.\frac{n_{h_2}/s_D}{n_\gamma/s_{\rm SM}}\right|_{T^{\rm cd}} \times \left.\frac{s_D}{s_{SM}}\right|_{T_{\rm dec}}\nonumber\\
&=& \frac{1}{2/g_{\rm SM}(T^{\rm cd})} \times \frac{2}{g_{\rm SM}(T_{\rm dec})} 
 = \frac{g_{\rm SM}(T^{\rm cd})}{g_{\rm SM}(T_{\rm dec})}\,.
\end{eqnarray}
Multiplying this result by an additional scaling factor $g_{\rm SM}(10\,{\rm GeV})/g_{\rm SM}(T^{\rm cd})$ we can obtain the value that may be compared with the bounds in Ref.~\cite{Hufnagel:2018bjp} 
\begin{eqnarray}
\left.\frac{n_{h_2}}{n_\gamma}\right|_{T^{\rm cd}=10\,{\rm GeV}} = \frac{g_{\rm SM}(10\,{\rm GeV})}{g_{\rm SM}(T_{\rm dec})} = 1.27\,,
\end{eqnarray}
where we have used $g_{\rm SM}(10\,{\rm GeV}) = 78.25$ and $g_{\rm SM}(T_{\rm dec} = 500~{\rm MeV})=61.75$. This ratio satisfies BBN and CMB constraints presented in the left panel of fig. 4 in Ref. \cite{Hufnagel:2018bjp}. Moreover, for shorter lifetimes $\tau_{h_2}<10$~s the bounds are weaker, therefore BBN and CMB bounds are always satisfied provided that the dark sector decouples before QCD phase transitions.

\begin{figure}[t!]
\centering
\includegraphics[width = 0.6\linewidth]{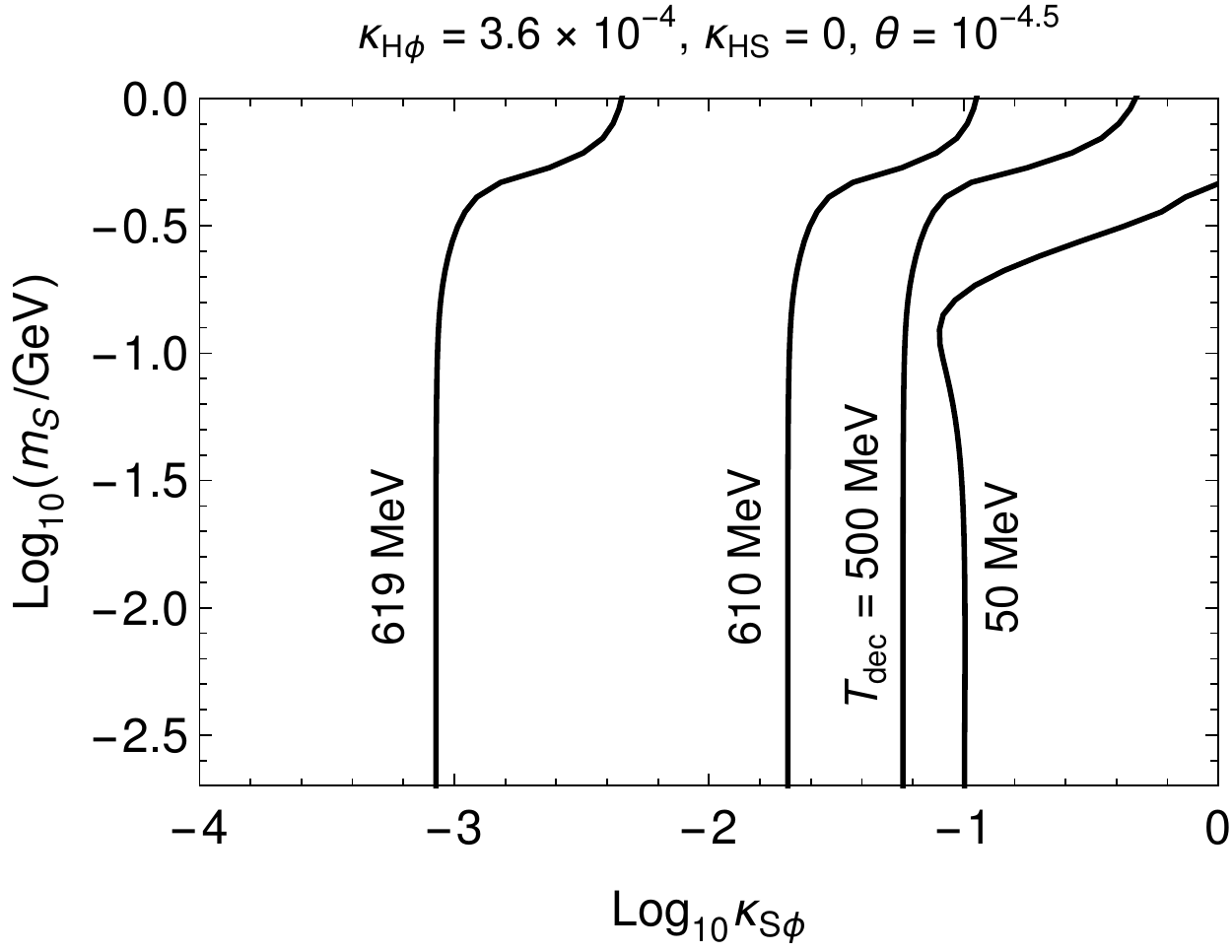}
\caption{The contours of the dark sector decoupling temperature $T_{\rm dec}$ in the $\kappa_{S\phi}$-$m_S$ plane choosing $\kappa_{H\phi} = 3.6\times 10^{-4}$ which maximizes the region $T_{\rm dec} > 500$~MeV for $\kappa_{HS}=0$ and $\theta = 10^{-4.5}$.}\label{decoupling}
\end{figure}

Sufficiently early decoupling of the dark sector can be achieved for suppressed values of the Higgs portal couplings. Note that the decoupling happens when the annihilation rate equals the Hubble rate $\Gamma(T_{\rm dec}) = H (T_{\rm dec})$~\cite{Cline:2013gha}. Below $T_{\rm dec}$, the annihilation cannot catch up with the cosmic expansion and it is effectively turned off. In the present model, there are two processes that set the decoupling temperature: $h_2 h_2 \to {\rm SM}\, {\rm SM}$ and $SS \to {\rm SM}\, {\rm SM}$. The annihilation rate of the first one scales with $\kappa^2_{H\phi}$, whereas for the second, the scaling is $\kappa^2_{S\phi}\theta^2/\kappa^2_{H\phi}$. Here, for simplicity, we assumed that $\kappa_{HS} \ll \kappa_{S\phi}, \kappa_{H\phi}$. We found that, for $s_\theta = 10^{-4.5}$ and $\kappa_{HS}=0$, the value $\kappa_{H\phi}=3.6\times 10^{-4}$ maximizes the region $T_{\rm dec}>500$~MeV in the plane $\kappa_{S\phi}$-$m_S$ of the parameter space~\footnote{It is possible to enlarge the region $T_{\rm dec}>500$~MeV by choosing the suitable nonzero value of $\kappa_{HS}$ of the same sign as $\kappa_{H\phi}\kappa_{S\phi}$, however the change is very mild. Moreover, the exact value of $\kappa_{HS}$ has moderate impact on the calculation of relic abundance and indirect detection bounds when S annihilates dominantly in $h_2h_2$ channel.}
In Fig.~\ref{decoupling}, we present the contours of decoupling temperature $T_{\rm dec}$ for this set of parameters. If $\kappa_{S\phi}\lesssim 0.05$, dark sector decoupling temperature does not vary much, but for larger values of $\kappa_{S\phi}$, it quickly drops below $500$~MeV. The contours of constant $T_{\rm dec}$ bend when $m_S$ approaches $1$~GeV increasing the allowed region, because for higher masses the density of $h_2$ becomes Boltzmann suppressed at the considered temperatures and therefore annihilation rates are diminished.

{In principle, also the SM-like Higgs decays into $SS$ and $h_2h_2$ are constrained by searches for invisible Higgs decays, but in the range of allowed mixing angles and suppressed Higgs portal couplings that we consider, these bounds are easily satisfied.}

\section{Dark Matter Direct Detection} 
\label{Sec_DD}

As mentioned before, in the strongly self-interacting fermionic DM model with an unstable ${\cal O}({\rm MeV})$ scalar mediator, the upper bound from the current DM direct detection experiments requires a too long mediator's lifetime that is not consistent with the BBN~\cite{Kainulainen:2015sva,Kaplinghat:2013yxa,Hufnagel:2018bjp}. 
This problem is one of the main motivations to study the present model with a stable scalar mediator.
Note that both $\chi$ and $S$ contribute to the observed DM relic density. Thus, their scatterings with nucleons would cause recoils observable in DM underground detectors, which could severely constrain the model. Note that both the scalar-nucleon $SN$ and fermion-nucleon $\chi N$ scatterings are mediated by the two Higgs particles $h_{1,2}$. As $h_2$ is assumed to be light, we need to take into account its light mediator effect in the DM-nucleon interactions~\cite{Li:2014vza,Geng:2016uqt}. 

\begin{figure}[]
\centering
\includegraphics[width = 0.7\linewidth]{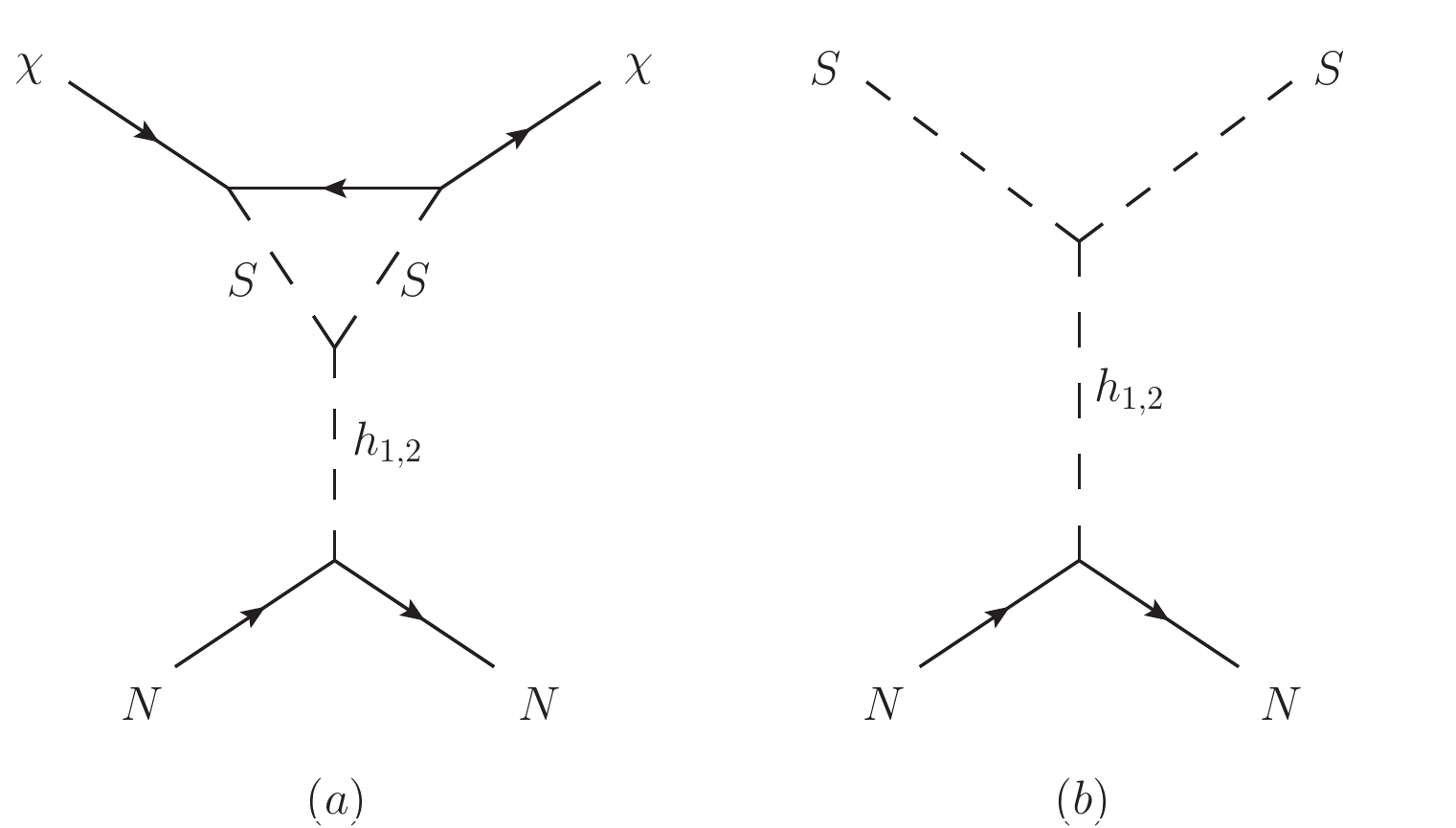}
\caption{The dominant Feynman diagrams for the direct searches for (a) the fermionic DM $\chi$ and (b) the stable light scalar mediator $S$. The Feynman diagram for $\bar{\chi}$ nuclear scatterings can be obtained by reversing the direction of the fermionic line in (a).}\label{dmdd}
\end{figure}

The scattering of the fermionic DM $\chi$ against nucleon is induced at one-loop order as shown in Fig.~\ref{dmdd}(a). The nuclear recoil cross section is given by
\begin{eqnarray}
\sigma_{\chi N} = \frac{g_Y^4 f_N^2}{256\pi^5} \frac{\mu_{\chi N}^2 m_N^2}{m_1^4 m_\chi^2} \frac{(\kappa_{S\phi}\kappa_{H\phi}-2\lambda_\phi \kappa_{HS})^2 v_\phi^4}{m_2^2 (4\mu^2_{\chi N}v^2+m_2^2)} F\left(\frac{m_S^2}{m_\chi^2}\right)^2\,,
\end{eqnarray}
with the effective Higgs-nucleon coupling $f_N \simeq 0.3$~\cite{Cline:2013gha,Alarcon:2011zs,Ling:2017jyz}. Here the factor $(4\mu^2_{\chi N}v^2 + m_2^2)$ in the denominator represents the possible light $h_2$ mediator effects~~\cite{Li:2014vza,Geng:2016uqt}, where $v \approx 220$~km/s denotes the typical DM velocity in the Milky Way relative to the Solar system, and the loop fucntion $F(t)$ is defined as follows
\begin{eqnarray}
F(t) &=& \int^1_0 dx \frac{x(1-x)}{x^2 + (1-x)t} \nonumber\\
& = & -\frac{t}{2[(4-t)t]^{3/2}} \Bigg\{(8-2t)\sqrt{t(4-t)} + (4-5t+t^2)\sqrt{t(4-t)}\ln t \nonumber\\
&&  - 2t(4-t)(3-t)\left({\rm ArcTan}\left[\sqrt{\frac{t}{4-t}}\right]-{\rm ArcTan}\left[\frac{t-2}{\sqrt{t(4-t)}}\right] \right) \Bigg\}\,,
\end{eqnarray}
which is valid when $0<t<4$. If we take the benchmark values for the model parameters as $g_Y\sim 0.1$, $m_\chi \sim 100$~GeV, and $m_S \sim 10$~MeV, the above formula gives the $\chi$-nucleon cross section to be of ${\cal O}(10^{-48}~{\rm cm}^2)$, which is well below the best experimental bound from XENON1T experiment~\cite{Aprile:2018dbl}. Hence, we can ignore the DM direct detection constraints on the heavy DM $\chi$ properties.

It turns out that the contribution from the stable mediator $S$ is also subdominant. Its tree-level scattering cross section on nucleon (see Fig.~\ref{dmdd}(b)) reads 
\begin{eqnarray}
\sigma_{SN} = \frac{f_N^2 (\kappa_{S\phi}\kappa_{H\phi}-2\lambda_\phi \kappa_{HS})^2}{4\pi} \frac{ \mu_{SN}^2 v_\phi^4 m_N^2}{m_1^4 m_2^4 m_S^2}\,,
\end{eqnarray}
with $\mu_{SN} \equiv m_S m_N/(m_S+m_N)$ being the reduced mass in the $S$-$N$ system. In the above formula we have ignored the momentum transfer for 
$h_1$ and even for $h_2$. Note that the mass of $S$ is expected to be of ${\cal O}(1 \sim 100~{\rm MeV})$. In this range, the nuclear recoil energy is well below the lowest thresholds obtained in the direct detection experiments. This problem can be overcome if one considers electron recoils \cite{Essig:2011nj,Essig:2017kqs}, however in the discussed model the Higgs-mediated DM-electron interactions are strongly suppressed. Furthermore, it is also possible to utilize the highly energetic mediators $S$ arising from $\chi$ annihilations \cite{Agashe:2014yua,McKeen:2018pbb} or collisions with cosmic rays \cite{Bringmann:2018cvk}, but these bounds are to weak to constrain the model.

Concluding, the current DM direct searches cannot put any useful constraints from either the heavy DM $\chi$ or the light stable mediator $S$ scattering.
Therefore we are able to avoid the conflict faced by the self-interacting fermionic DM models with the unstable mediator investigated in Refs.~\cite{Kainulainen:2015sva,Kaplinghat:2013yxa,Hufnagel:2018bjp}. 

%%%%%%%%%%%%%%%%%%%%%%%%%%%%%%%%%%%%%%%%%%%%%%%%%%%%%%%%%%%%%%%%%%%%%%%%%%%%%%%%%%%%%%%%%
\section{Dark Matter Indirect Detections}
\label{Sec_DMID}
It is well known that one of the most relevant difficulties to construct a viable model with sufficient DM self-interactions originates from DM indirect detection constraints~\cite{Feng:2010gw}. This is because almost the same intermediate DM bound states that generate the DM self-interactions would also induce the Sommerfeld enhancement of the DM annihilation~\cite{Bringmann:2016din,Cirelli:2016rnw}, which is the main target for DM indirect searches. At present, the most important DM indirect detection constraints comes from the Planck measurement of CMB spectral distortion~\cite{Ade:2015xua}, a gamma-ray probe of the dwarf spheroidal galaxies by Fermi-LAT~\cite{Ackermann:2015zua}, and the AMS-02 search of high energy positrons in the Milky Way~\cite{AMS1, AMS2}. As shown in Refs.~\cite{Ahmed:2017dbb, Duerr:2018mbd}, if the light vector mediator which helps to form the DM bound states is stable, such constraints can be efficiently avoided, since most annihilations of dominant DM particles go into invisible light mediators. A similar mechanism takes place in the present model with a stable light scalar mediator, since the dominant fermionic DM density is provided by the freeze-out of its unobservable annihilation channel $\chi \bar{\chi} \to S S$. Nevertheless, there are still several channels in the dark sector which could be probed by the indirect detections.

One important channel is the semi-annihilation process $\chi\chi (\bar{\chi}\bar{\chi}) \to S h_{1,2}$ followed by the $h_{1,2}$ decays. However, this process can be proven to be dominated by the $p$-wave cross section, which approaches zero in the non-relativistic limit with the relative velocity $v_{\rm rel} \to 0$, in spite of a potential significant $p$-wave Sommerfeld enhancement. We have explicitly computed the cross sections with the DM velocities of dwarf galaxies and the Milky Way, and compared the results with the constraints given by AMS-02 and Fermi-LAT~\cite{Elor:2015bho}. As a result, this process cannot impose any relevant constraints (similarly to the unstable mediator case ~\cite{Kahlhoefer:2017umn}). This conclusion may be altered, when one considers $2\rightarrow 3$ process with extra initial state radiation of the mediator which in some cases occur to be s-wave dominated \cite{Bell:2017irk}. However, for the fermionic DM and purely scalar mediator, it remains p-wave (contrary to e.g. purely pseudoscalar mediator) \cite{Bell:2017irk,Kahlhoefer:2017umn}.

Another possible signal stems from the stable scalar mediator $S$ annihilation into the visible particles via the Higgs portal interactions. {The dominant annihilation channel for $S$ is the pair of $h_2$ which subsequently decay into the electrons and positrons}. As long as $S$ is assumed to be lighter than $1$~GeV, the relevant constraint can only be given by the CMB observation, while the energy of produced $e^+/e^-$ is always too small to be constrained by Fermi-LAT or AMS-02. The explicit upper bound from CMB is given as follows
\begin{eqnarray}
\langle \sigma v \rangle_{SS\to h_2 h_2}  \left(\frac{\Omega_S h^2}{\Omega_{\rm DM} h^2}\right)^2 < \langle\sigma v_{\rm rel}\rangle^{4e^{\pm}}_{\rm CMB}(m_S)\,,
\end{eqnarray}
where $\langle \sigma v_{\rm rel} \rangle^{4 e^\pm}_{\rm CMB} (m_S)$ represents the upper bound given in Ref.~\cite{Slatyer:2015jla} on the $S$ annihilation cross section with two pairs of electrons and positrons in the final state. The factor $(\Omega_S h^2 / \Omega_{\rm DM} h^2)^2$ accounts for the suppression from the $S$ energy density fraction, with $\Omega_{\rm DM} h^2 \simeq 0.12$ the total measured DM relic density~\cite{PDG}.   

Moreover, the model is further constrained by the DM indirect searches for the process $\chi S \to \bar{\chi} h_2$ and $\bar{\chi} S \to \chi h_2$, in which the signal also originates from the $h_2$ decays. For simplicity, we will use $\chi S \to \bar{\chi} h_2$ to denote both processes in the following formulas and plots. In the non-relativistic limit relevant to the DM indirect searches, the energy of the produced $h_2$ can be estimated to be the mass of $S$. By considering the suppression from respective density fractions of the fermionic DM and the stable mediator, we can write down the following constraint for this process
\begin{eqnarray}
\frac{1}{2} \langle \sigma v \rangle_{\chi S \to \bar{\chi} h_2} \left(\frac{\Omega_S h^2}{\Omega_{\rm DM} h^2}\right) \left(\frac{\Omega_\chi h^2}{\Omega_{\rm DM} h^2}\right) < \langle\sigma v_{\rm rel}\rangle^{4e^{\pm}}_{\rm CMB}(m_S)\,,
\end{eqnarray}   
where the factor $1/2$ on the left hand side accounts for the fact that only one pair of $e^\pm$ is generated in this reaction.

Note that in contrast to the annihilation of fermionic DM particles, the above two processes are not subject to the Sommerfeld enhancement, and the corresponding cross sections are $s$-wave dominated so that they approach non-zero values at the non-relativstic limit $v_{\rm rel} \to 0$.

%%%%%%%%%%%%%%%%%%%%%%%%%%%%%%%%%%%%%%%%%%%%%%%%%%%%%%%%%%%%%%%%%
\section{Dark Matter Self-Interactions}
\label{Sec_DMSI}
The main motivation for the introduction of the DM self-interactions is to solve the small-scale structure problems in our Universe, such as the cusp-vs-core problem and the too-big-to-fail problem at the scale of the dwarf galaxies. By fitting the data, the required DM self-interaction per unit DM mass should be $0.1~{\rm cm}^2/{\rm g}<\sigma_T/m_\chi < 10~{\rm cm}^2/{\rm g}$ for dwarf galaxies with the typical DM velocity at $v \simeq 30$~km/s~\cite{deLaix:1995vi,Spergel:1999mh,Vogelsberger:2012ku,Zavala:2012us,Rocha:2012jg,Peter:2012jh, Kaplinghat:2015aga,Tulin:2017ara}, where $\sigma_T$ denotes the momentum transfer cross sections in the DM scatterings. On the other hand, such a strong DM self-scattering can also leads to the observable effects at the galaxy cluster scale with $v\simeq 1000$~km/s. The absence of these effects in the the galaxy cluster data strongly constrains the DM self-interactions at this cosmological scale, with the conservative upper limit given by $\sigma_T/m_\chi < 1~{\rm cm}^2/{\rm g}$~\cite{Clowe:2003tk,Markevitch:2003at,Randall:2007ph,Harvey:2015hha,Kahlhoefer:2013dca}. 

With the mass hierarchy between the scalar $S$ and the Dirac fermion $\chi$, the DM self-interactions can be greatly enhanced by the formation of the $\chi\chi$ or $\chi\bar{\chi}$ bound states via the mediation of $S$~\cite{Buckley:2009in,Feng:2009mn,Tulin:2012wi,Tulin:2013teo,Tulin:2012wi,Tulin:2013teo}. Since the incoming and outgoing states are different, it is difficult in analyzing the DM self-interactions in this form of the Yukawa couplings. In order to overcome this problem, we define the following two Majorana fermions
\begin{eqnarray}
\chi_+ = \frac{1}{\sqrt{2}} (\chi+\chi^c)\,,\quad \quad \chi_- = \frac{i}{\sqrt{2}}(\chi-\chi^c)\,.
\end{eqnarray}
With this transformation of fermion basis, the corresponding Lagrangian is modified to
\begin{eqnarray}\label{LagNew}
{\cal L}_\chi = i\bar{\chi}_+ \slashed{\partial}\chi_+ + i\bar{\chi}_- \slashed{\partial}\chi_- - \frac{m_\chi}{2}(\bar{\chi}_+ \chi_+ + \bar{\chi}_-\chi_-) - \frac{g_Y}{2}S(\bar{\chi}_+\chi_+ - \bar{\chi}_- \chi_-)\,.
\end{eqnarray}
Accordingly, the original stabilizing symmetries are transformed into the following three $Z_2$ ones
\begin{eqnarray}
&Z_2^1:& \quad \quad S \to -S\,,\quad \chi_+ \to \chi_-\,;\nonumber\\
&Z_2^2:& \quad \quad \chi_+ \to \chi_+\,;\nonumber\\
&Z_2^3:& \quad \quad \chi_- \to \chi_-\,.
\end{eqnarray} 
Thus, the scalar $S$ and two Majorana fermions $\chi_\pm$ are all DM candidates. With the Lagrangian in Eq.~(\ref{LagNew}), the Yukawa terms now are diagonalized and the incoming and outgoing fermions keep same, which simplifies our following discussion. In particular, it is easy to see that the potential in the $\chi_+ \chi_+$ and $\chi_- \chi_-$ systems are attractive, given by
\begin{eqnarray}\label{Va}
V_{a} = -\frac{g^2}{4\pi r} e^{-m_S r}\,,
\end{eqnarray}
while $\chi_+ \chi_-$ scatterings are repulsive with the following potential:
\begin{eqnarray}\label{Vr}
V_{r} = +\frac{g^2}{4\pi r} e^{-m_S r}\,.
\end{eqnarray}

\begin{figure}[]
\centering
\includegraphics[width = 0.6\linewidth]{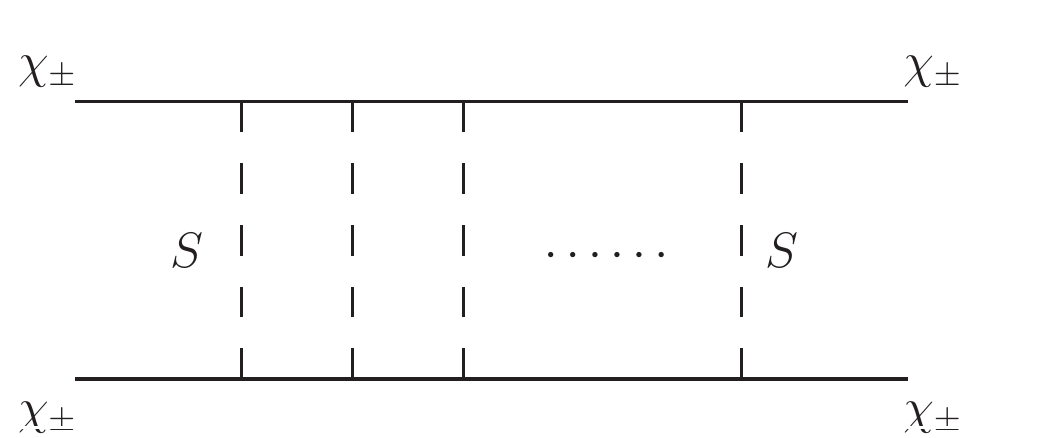}
\caption{Pictorial illustration of the multiple exchange of the stable mediator $S$ to generate the self-interactions between the fermionic DM $\chi_{\pm}$. }
\label{dmsi}
\end{figure}
Accordingly, the momentum transfer cross section $\sigma_T$, which characterizes the DM self-interactions, is defined in the present model as follows~\cite{Kahlhoefer:2017umn}
\begin{eqnarray}
\sigma_T \equiv \frac{1}{4} \left( \sigma_T^{++} + \sigma_T^{--} + \sigma_T^{+-} + \sigma_T^{-+}\right) = \frac{1}{2} \left( \sigma_T^{a} + \sigma_T^{r} \right)\,,
\end{eqnarray}
with 
\begin{eqnarray}
\sigma_T^{a,r} \equiv 2\pi \int^1_{-1} \left(\frac{d\sigma}{d\Omega}\right)^{a,r} (1-|\cos\theta |) d \cos\theta\,,
\end{eqnarray}
where $\sigma_T^{\pm\pm}$ and $\sigma_T^{\pm\mp}$ are the cross sections for the $\chi_\pm \chi_\pm$  and $\chi_\pm \chi_\mp$ scatterings, respectively, and, as argued before, correspond to the ones induced by the attractive and repulsive potentials, {\it i.e.}, $\sigma_T^a$ and $\sigma_T^r$. In our computation of the momentum transfer cross sections, we follow Ref.~\cite{Tulin:2013teo,Kahlhoefer:2017umn} to numerically solve the Schr$\ddot{\rm o}$dinger equations with the potentials given in Eqs.~(\ref{Va}) and (\ref{Vr}), which effectively sums over multiple exchanges of $S$ in the interactions between two heavy fermionic DM particles as illustrated in Fig.~\ref{dmsi}. This is particularly useful when we work in the parameter space of the so-called quantum resonant regime in which the attractive potential leads to the formation of resonances of $\chi$ bound states. In our calculation, we also take into account effects arising from the quantum indistinguishability of identical particles in the initial and final states following the procedure given in Ref.~\cite{Kahlhoefer:2017umn}.

Note that when $g^2 m_\chi/(4\pi m_S)\ll 1$, the non-perturbative momentum transfer cross sections would reduce to the analytical formula given in Ref.~\cite{Kahlhoefer:2017umn} for the Born approximation. In another limit where $m_\chi v/m_S >10$, we are entering the classical regime in which it is very difficult to solve the Schr$\ddot{\rm o}$dinger equations and the approximate expressions for the momentum transfer cross sections given in Ref.~\cite{Cyr-Racine:2015ihg} can be applied. In our following numerical calculations, we shall adopt both analytical expressions in these two parameter regimes.

%%%%%%%%%%%%%%%%%%%%%%%%%%%%%%%%%%%%%%%%%%%%%%%%%%%%%%%%%%%%%%%%%%%%%%%%%%%%%%%%%%%%%%%%%
\section{Numerical Results}
\label{Sec_NumRes}
\begin{figure}[]
\centering
\includegraphics[width = 0.48 \linewidth]{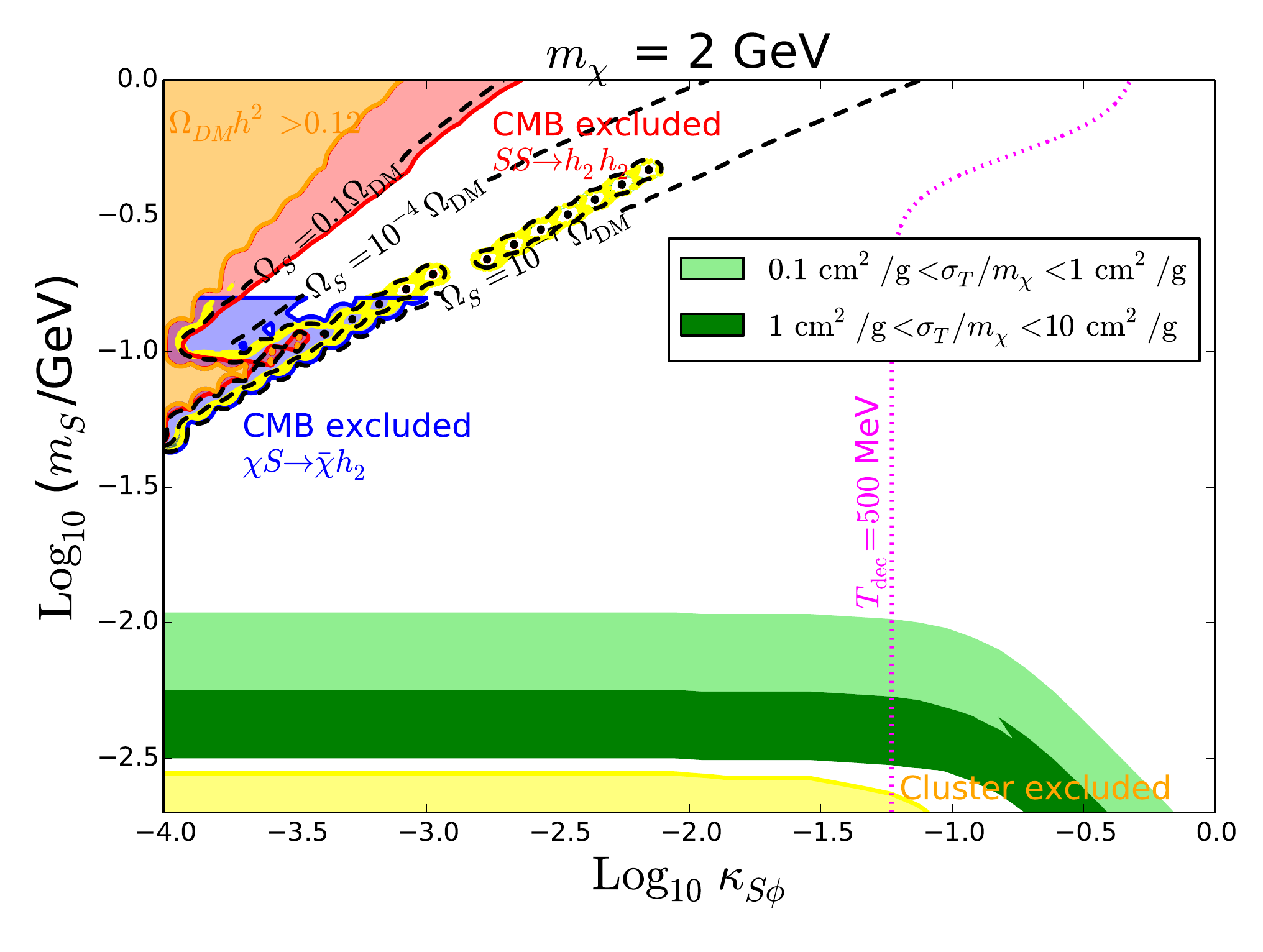}
\includegraphics[width = 0.48 \linewidth]{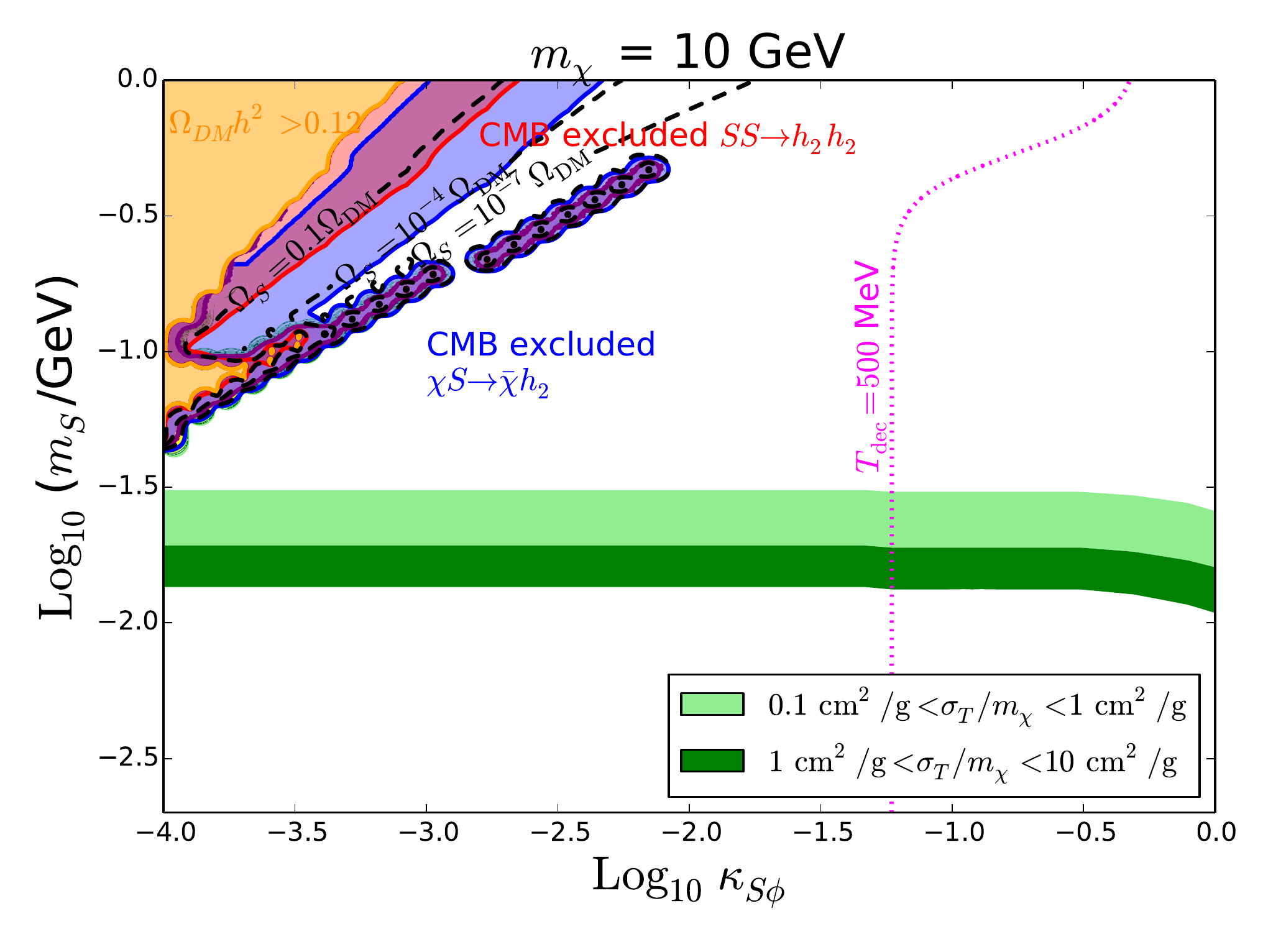}
\includegraphics[width = 0.48 \linewidth]{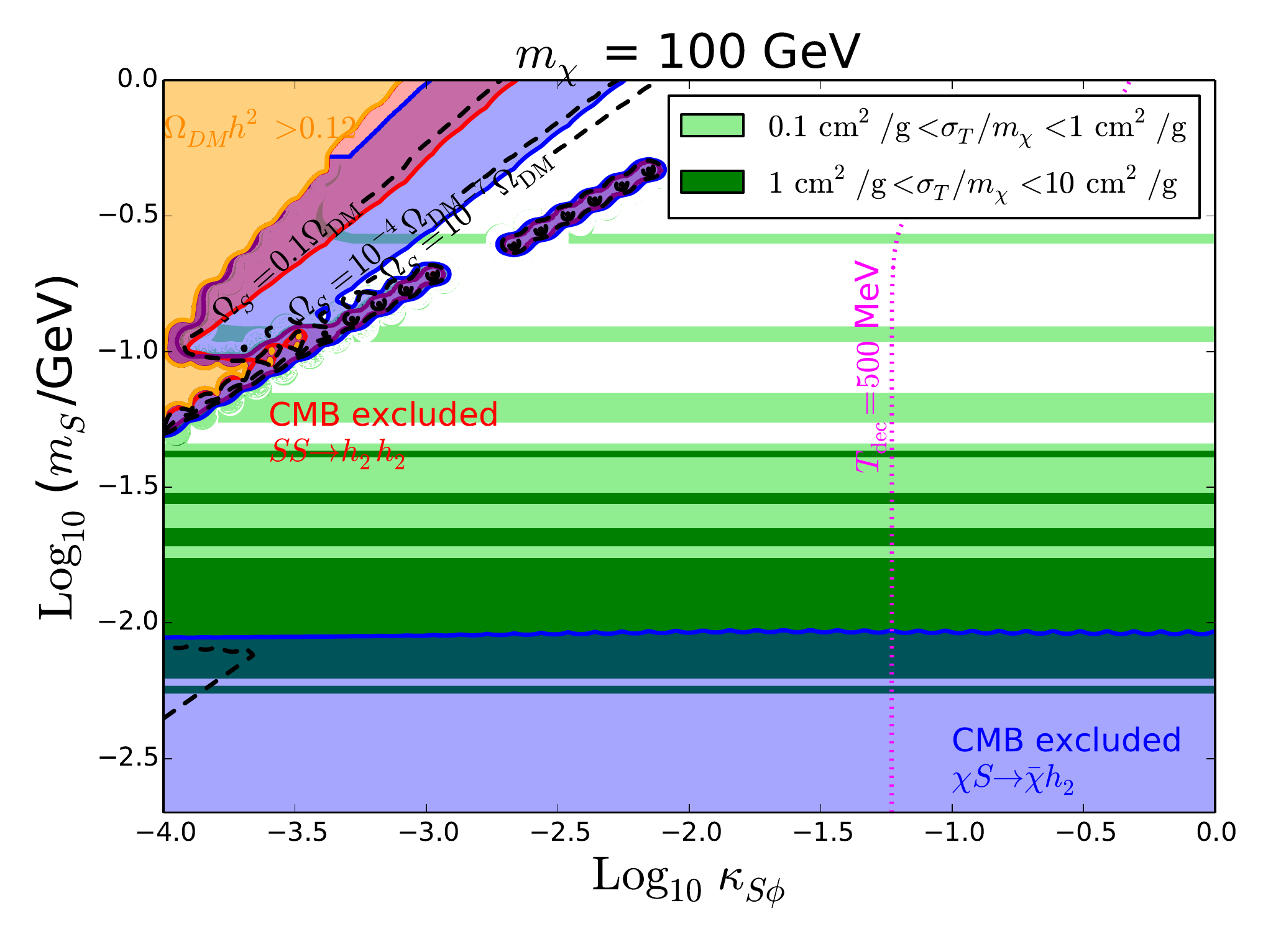}
\includegraphics[width = 0.48 \linewidth]{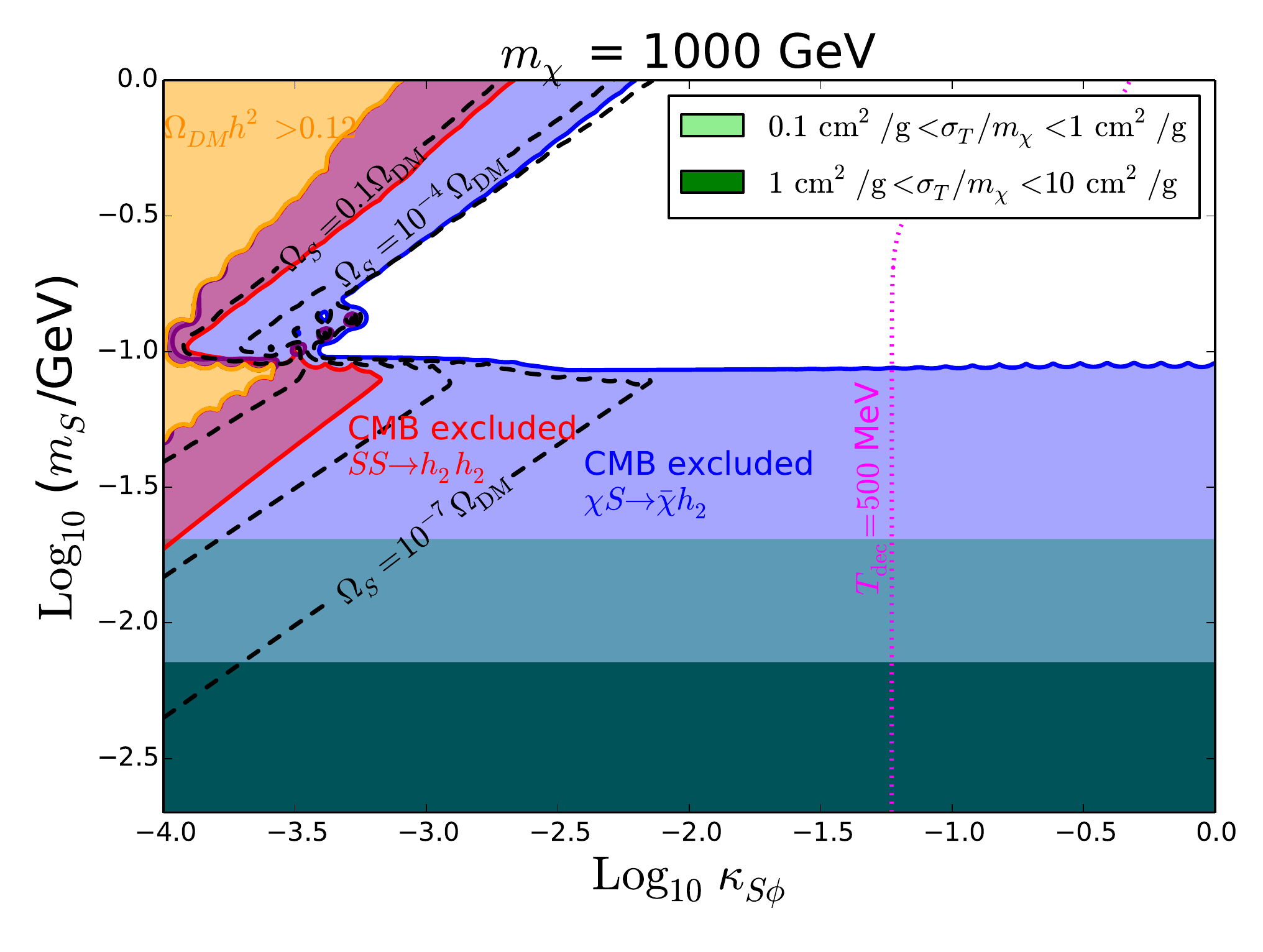}
\caption{Constraints in the $m_{S}$-$m_\chi$ plane of the parameter space in the regions with substantial DM self-interactions for $m_\chi = 2$~GeV (upper left panel), $m_\chi = 10$~GeV (upper right panel), $m_\chi = 100$~GeV (lower left panel) and $m_\chi = 1000$~GeV (lower right panel). In each panel the second (dark) Higgs boson mass is fixed to be $m_2 = 1.5$~MeV, while the dark Yukawa coupling $g_Y$ is obtained by requiring $\Omega_\chi h^2 + \Omega_S h^2 \simeq 0.12$ at each point. The orange shaded areas are excluded, because DM relic density exceeds the measured value for any chosen $g_Y$. The light and dark green regions represent the parameter space which can generate a momentum transfer cross section $\sigma_T$ of the fermionic DM self-interaction at the scale of dwarf galaxies in the range of $0.1\,{\rm cm^2/g}< \sigma_T/m_\chi < 1\,{\rm cm^2/g}$ and $1\,{\rm cm^2/g}< \sigma_T/m_\chi < 10\,{\rm cm^2/g}$, respectively, while the yellow region is excluded by the bound $\sigma_T/m_\chi \lesssim 1~{\rm cm^2/g}$ on the scale of galaxy clusters. The red and blue shaded regions indicate parameters which are ruled out by the CMB constraints on the late energy injections from the processes $SS \to h_2 h_2$ and $\chi S \to \bar{\chi} h_2$ ($\bar{\chi} S \to \chi h_2$). To the left of the magenta dotted curve $T_{\rm dec} = 500$~MeV, dark sector fully decouples before QCD phase transition.}\label{mp}
\end{figure}

In this section, we present our numerical results in relevant regions of the parameter space. We begin by presenting the results in Fig.~\ref{mp} spanned by the portal coupling $\kappa_{S\phi}$ and the light stable mediator  mass $m_S$ for different values of the fermionic DM mass $m_\chi = 2$, 10, 100, 1000~GeV from top left to bottom right, respectively.  
Following the discussion in Sec.~\ref{Sec_h2}, we have fixed the mass of the dark Higgs boson $h_2$ to be $m_{h_2} = 1.5$~MeV to avoid the CMB and BBN constraints for the late-time $h_2$ decays as much as possible. Note that to make the annihilation channel $S S \to h_2 h_2$ kinematically allowed in order to deplete the mass density of $S$, we choose the lower limit of $S$ mass to be $m_S = 2$~MeV in all panels of Fig.~\ref{mp}. Then we fix $\kappa_{H\phi} = 3.6\times 10^{-4}$, $\kappa_{HS} = 0.0$, $s_\theta= 10^{-4.5}$, so that we are left with only four free parameters. Therefore, the Yukawa coupling $g_Y$ in Eq.~(\ref{YchiS}) can be determined for each set of given values of ($m_\chi$, $m_S$, $\kappa_{S\phi}$) by the requirement that $\chi$ and $S$ constitute all of the measured DM relic density with $\Omega_{\rm DM} h^2 = \Omega_{\chi}h^2 + \Omega_{S} h^2 \simeq 0.12$. We also show,  in Fig.~\ref{mp}, the contours of constant relic density fraction $\Omega_S/\Omega_{\rm DM}$ as black dashed lines with the three curves in each panel corresponding to $\Omega_S/\Omega_{\rm DM} = 0.1$, $10^{-4}$, and $10^{-7}$, from left to right respectively. In Fig.~\ref{mp}, this fraction is found to rise with decreasing portal coupling $\kappa_{S\phi}$ and increasing $m_S$, which implies that the cross section of $SS \to h_2 h_2$ process is reduced. At some point, the cross section of $SS \to h_2 h_2$ becomes too small to eliminate enough $S$ so that, no matter what value of $g$ we choose, the relic density of $S$ alone would overclose the Universe. This part of parameter space should be excluded as the orange shaded regions indicate. Note that in all of these four panels it is seen that $S$ abundance increases significantly around the curve $m_S = \sqrt{\kappa_{S\phi}}v_\phi$. This feature can be understood as follows. It can be proven that, for the parameter space of interest, the $SS \to h_2 h_2$ amplitude is dominated by the Feynman diagrams ($d_1$) and $(d_3)$ in Fig.~\ref{dmrd} with the corresponding amplitudes given below:
\begin{eqnarray}\label{cancel}
i{\cal M}_{(d_1)} &\approx & i c_\theta^2 \kappa_{S\phi}^2 \frac{v_\phi^2}{m_S^2}\,,\nonumber\\
i{\cal M}_{(d_3)} &\approx & -i c_\theta^2 \kappa_{S\phi}\,.
\end{eqnarray} 
It is easy to see that when the relation $m_S = \sqrt{\kappa_{S\phi}}v_\phi$ holds, the total amplitude $i{\cal M}_{(d)} = i{\cal M}_{(d_1)} + i{\cal M}_{(d_3)} $ approaches to zero. It is this amplitude cancellation that causes the severe reduction of the $SS\to h_2 h_2$ cross section and thus the large $S$ relic density.

The main purpose of this paper is to provide a viable DM model in which the strong DM self-interactions can resolve the structure problems at the dwarf galaxy scale. In Fig.~\ref{mp}, we show the light and dark green regions of the parameter space where the momentum transfer cross section of $\chi$ is generated in the ranges of $0.1\,{\rm cm^2/g} < \sigma_T/m_\chi < 1\,{\rm cm^2/g}$ and $1\,{\rm cm^2/g} < \sigma_T/m_\chi < 10\,{\rm cm^2/g}$, respectively. We also display the bound on the DM self-interactions $\sigma_T/m_\chi \lesssim 1\,{\rm cm^2/g}$ from the galaxy cluster scale as the yellow shaded area. Note that this constraint is relevant only when the dominant DM particle $\chi$ is as light as $m_\chi = 2$~GeV. For heavier $\chi$, it is always satisfied for the whole parameter regions of interest. Moreover, when $m_\chi = 100$~GeV, the region which can explain the cosmological small-scale problems appears in the form of many spikes that are a characteristic feature of the quantum resonant regime discussed in Sec.~\ref{Sec_DMSI}. On the other hand, the relevant parameter regions for light fermionic DM masses $m_\chi = 2$ and 10 GeV correspond to the Born regime, while when $\chi$ becomes as heavy as ${\cal O}(1000~{\rm GeV})$, we enter the non-perturbative classical regime to generate large enough DM self-interactions. 

Furthermore, it is interesting to see that in all of the panels of Fig.~\ref{mp}, most of the signal regions with $\Omega_S \lesssim 0.1 \Omega_{\rm DM}$ are represented as straight bands, indicating that the DM self-interaction cross section is insensitive to $\kappa_{S\phi}$. This fact can be understood as follows. In such regions with $\Omega_S \lesssim 0.1 \Omega_{\rm DM}$, the observed DM relic density is determined by that of the fermionic DM, with its annihilations dominated by $\chi\bar{\chi} \to SS$. Thus, with a fixed fermionic DM mass in each plot, the Yukawa coupling $g_Y$ remains invariant as a consequence of a nearly constant $\langle \sigma v\rangle_{\chi\bar{\chi} \to SS}$, so the fermionic DM self-interaction is only the function of $m_S$. There are, however, two exceptions. The first case corresponds to the parameter regions with $\Omega_S \gtrsim 0.1 \Omega_{\rm DM}$ in the lower left panel of Fig.~\ref{mp}, in which the signal bands turn upward. The explanation is, when $S$ occupies a fraction $\gtrsim 0.1$ of the DM relic abundance, the fermionic DM abundance should be reduced and the value of $g_Y$ varies accordingly. The second exception is the region where $\kappa_{S\phi}$ is large enough and the bands turn down into lower $S$ masses as shown in the plots with $m_\chi = 2$~GeV and 10~GeV. This is just the reflection of the fact that, along with the increase of $\kappa_{S\phi}$, the semi-annihilation channel $\chi\chi\to S h_2$ plays a more and more important role in the decoupling of the dominant DM component $\chi$.

We also show the constraints from DM indirect detection in Fig.~\ref{mp}. As discussed in Sec.~\ref{Sec_DMID}, the constraints originate mainly from the CMB upper bounds on energy injections during the recombination era due to the processes $SS \to h_2 h_2$ and $\chi S \to \bar{\chi} h_2$ ($\bar{\chi} S \to \chi h_2$), which are represented as red and blue shaded areas in Fig.~\ref{mp}. From these four plots, it is easy to see that the annihilation of $SS$ into a pair of $h_2$ strongly constrains the parameter space of a large $S$ relic density fraction $\Omega_{S}/\Omega_{\rm DM} > 0.1$. On the other hand, the energy injection induced by the process $\chi S \to \bar{\chi} h_2$ can exclude additional regions. Firstly, it is found that the excluded region in the left top plot of Fig.~\ref{mp} have an upper boundary, which reflect the fact that the cross section of this process is suppressed too much when $m_S$ goes up beyond the boundary. Secondly, below this $S$ mass limit in the plot with $m_\chi = 2$~GeV and in other three plots, this process excludes more parameter space than $SS \to h_2 h_2$, extending the exclusion region deeply into the part with even smaller $S$ density fraction. This can be understood as follows: the scattering rate between $S$ and $\chi$($\bar{\chi}$) is proportional only to the first power of the density fraction of $S$, while the rate for the $S$ self-annihilation is further suppressed by the two powers. Lastly, when the mass of $\chi$ becomes to be of ${\cal O}(100\sim 1000~{\rm GeV})$, the region with lighter $S$ is excluded by the $\chi S \to \bar{\chi} h_2$ signal, which is the result of large coupling $\kappa_{S\phi}$ that overcompensate the suppressed $S$ density when computing the rate for this process.

{
The magenta dotted curve in Fig.~\ref{mp} encloses the region where dark sector decouples at temperatures larger than $T_{\rm dec} = 500$~MeV. Therefore, only for smaller values of $\kappa_S\phi$ dark Higgs fully decouples before QCD phase transition and, as explained in Sec.~\ref{Sec_h2}, the BBN and CMB bounds are evaded. }

It is clearly seen from Fig.~\ref{mp} that, when the dominant fermionic DM mass is below several hundred GeV, there is always a large parameter space which can accommodate DM relic density and explain the cosmological small-scale problems while satisfying the constraints of DM indirect detections. For the remaining parameter regions, the DM density fraction from the stable light mediator $S$ is always subdominant, which is constrained by the late-time electromagnetic energy injection during the recombination from the processes $SS \to h_2 h_2$ and $\chi S \to \bar{\chi} h_2$. However, in the bottom right panel of Fig.~\ref{mp}, the required DM self-interaction contours are excluded by the above two processes during the CMB formation, therefore this model disfavors the fermionic DM candidate as heavy as several TeV. 

\begin{figure}[ht]
\centering
\includegraphics[width = 0.48 \linewidth]{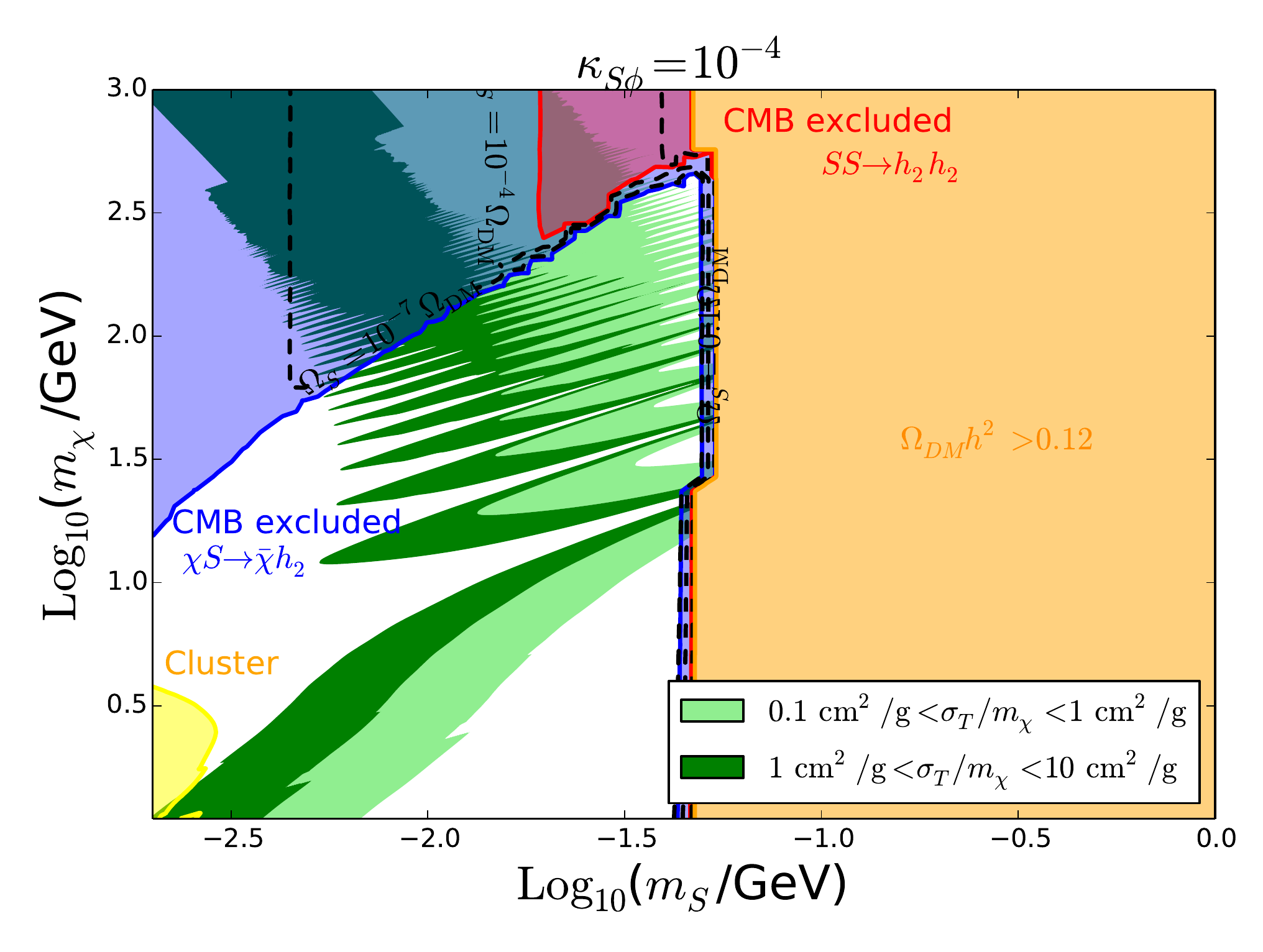}
\includegraphics[width = 0.48 \linewidth]{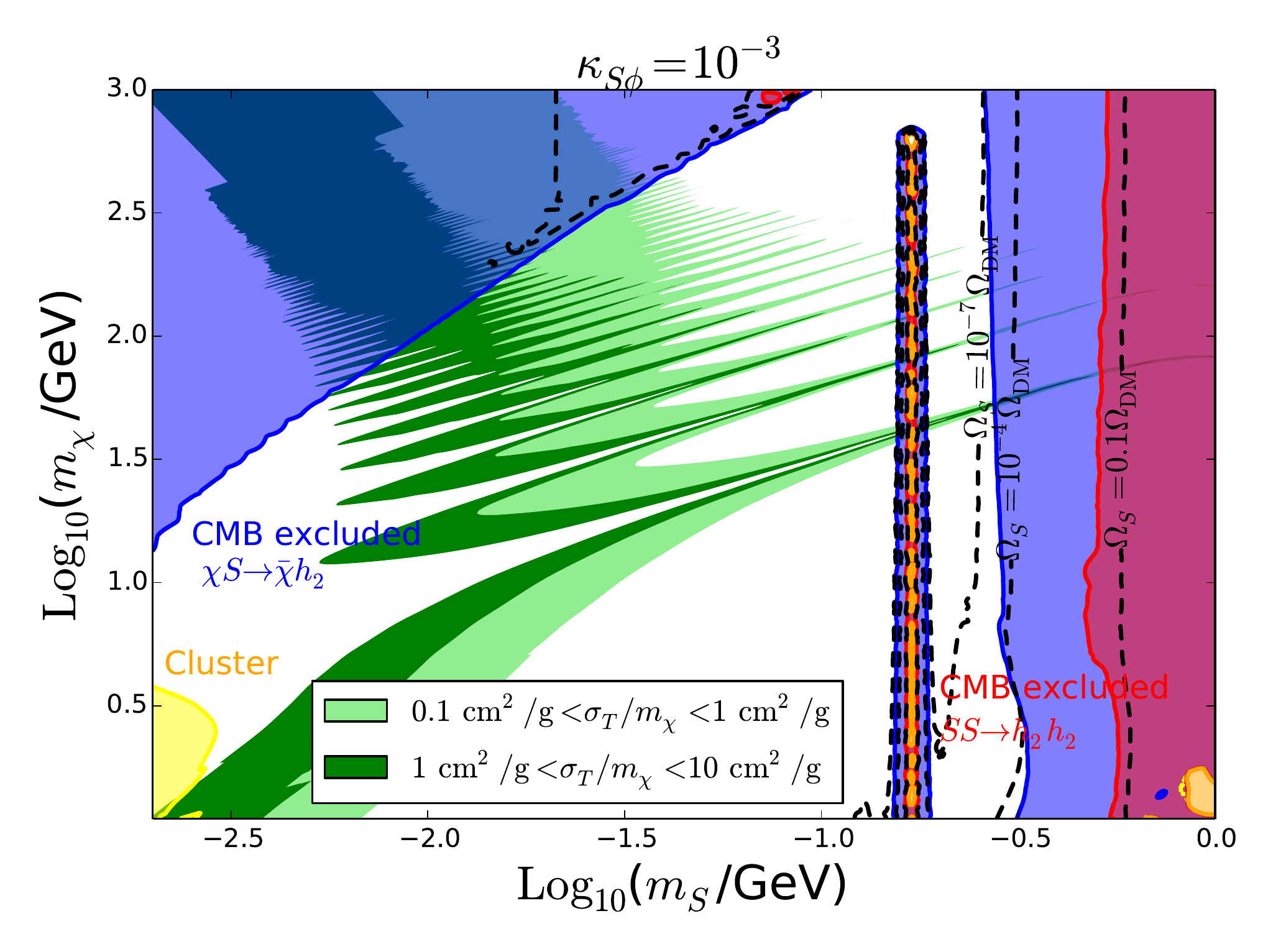}
\includegraphics[width = 0.48 \linewidth]{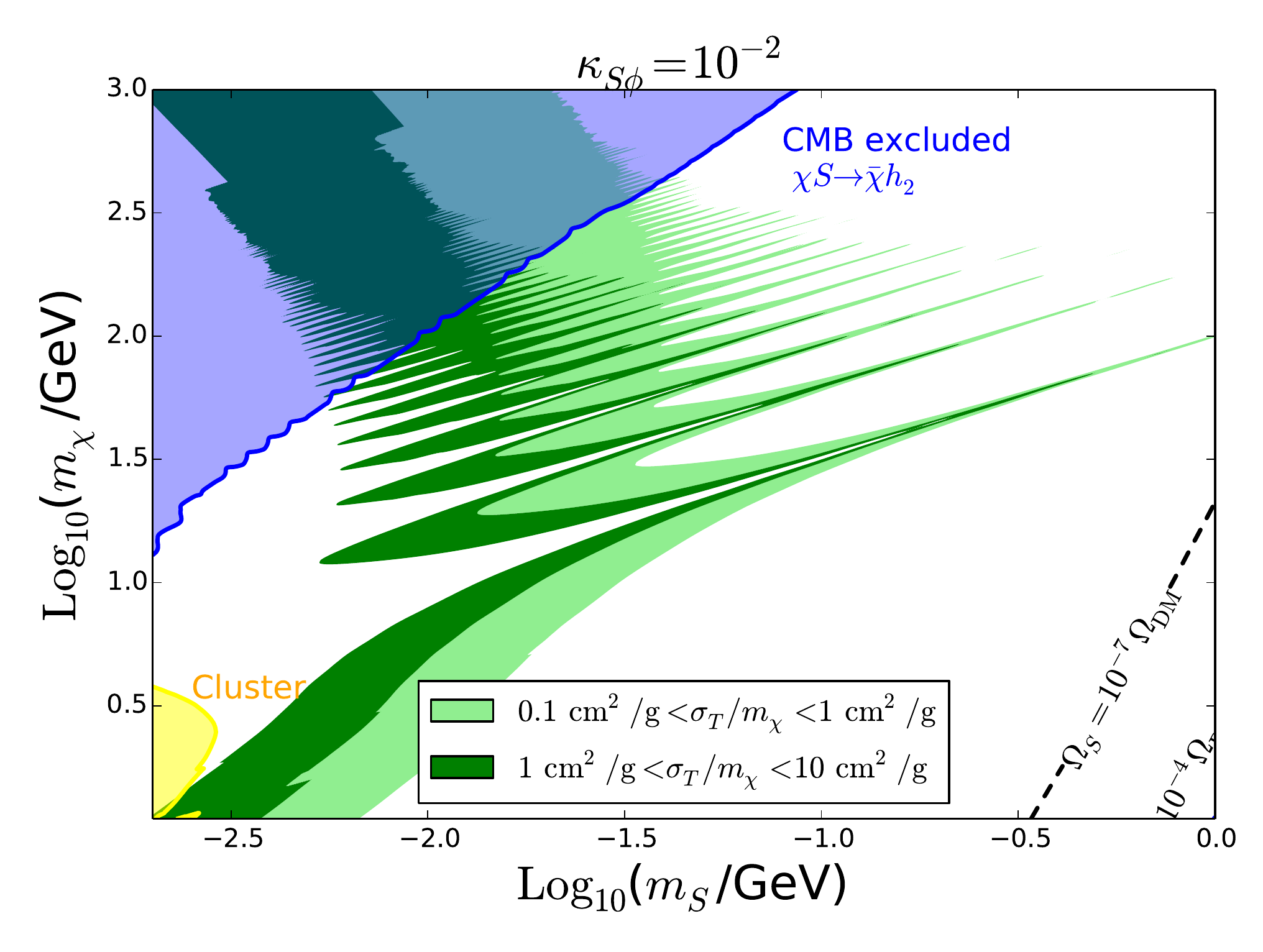}
\includegraphics[width = 0.48 \linewidth]{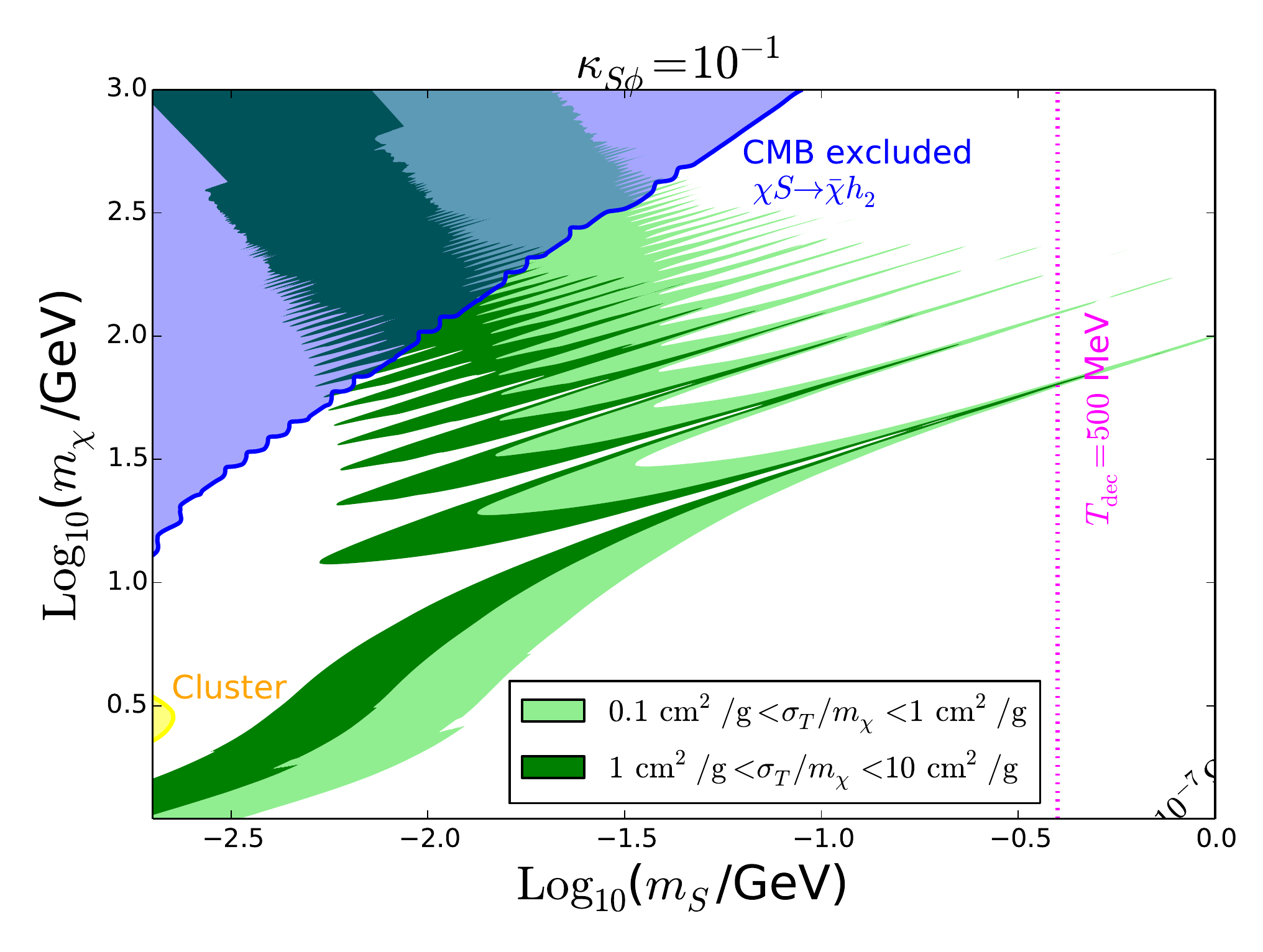}
\caption{Constraints in the $m_{S}$-$m_\chi$ plane of the parameter space in the regions with substantial DM self-interactions for $\kappa_{S\phi} = 10^{-4}$ (upper left panel), $\kappa_{S\phi} = 10^{-3}$ (upper right panel), $\kappa_{S\phi} = 10^{-2}$ (lower left panel) and $\kappa_{S\phi} = 10^{-1}$ (lower right panel). The color codings of each panel are the same as in Fig.~\ref{mp}.}\label{kap}
\end{figure}
We also present the same data points in the $m_S$-$m_\chi$ planes in Fig.~\ref{kap}, with the four panels corresponding to the cases with $\kappa_{S\phi}= 10^{-4}$, $10^{-3}$, $10^{-2}$, and $10^{-1}$ from upper left to lower right. The color coding in each panel is the same as that in Fig.~\ref{mp}. In each of the four panels of Fig.~\ref{kap}, a significant fraction of strongly DM self-interaction parameter spaces is allowed by current DM constraints. In particular, from the upper left panel with $\kappa_{S\phi} = 10^{-4}$, it follows that the combinations of constraints from DM self-interactions at galaxy clusters and from CMB observations constrain the allowed region to the corner with a fermionic DM mass $m_\chi \lesssim 300$~GeV and a small mediator mass $m_S < 40$~MeV. On the other hand, with the increase of $\kappa_{S\phi}$, the upper bounds from energy injections at recombination becomes more and more relaxed, and larger part of the parameter space is allowed to account for the small-scale structure problems. It is seen from Fig.~\ref{kap} that the CMB constraint from $SS\to h_2 h_2$ is sensitive to the $S$ density fraction and excludes at least regions with $\Omega_S/\Omega_{DM} > 0.1$, while the process $\chi S \to \bar{\chi} h_2$ implies a strong limit for the region with large fermionic DM mass $m_\chi$. Especially, the latter excludes all DM self-interaction signal regions with $m_\chi > 1$~TeV, which is consistent with the results shown in Fig.~\ref{mp}. Furthermore, note that the $S$ relic abundance increases abruptly around $m_S = 177$~MeV in the plot with $\kappa_{S\phi} = 10^{-3}$, which is just the manifestation of the cancellation of the $SS\to h_2 h_2$ amplitude discussed at the beginning of this section. {Finally, for $\kappa_{S\phi} = 0.1$ the coupling between dark and visible sectors is stronger and only to the right of the magenta dotted curve $T_{\rm dec}=500$~MeV the dark Higgs decay is in agreement with the BBN and CMB constraints.}

%However, along the same line of thinking, we hope to have the same peak feature at $m_S = 560$~MeV for the plot with $\kappa_{S\phi} = 10^{-2}$ }, which is absent in our scanning. The reason is, when $\kappa_{S\phi}$ is as large as $10^{-2}$ and the cancellation occurs for $SS \to h_2 h_2$, the conversion process can reduce the $S$ relic abundance to the acceptable level.

%%%%%%%%%%%%%%%%%%%%%%%%%%%%%%%%%%%%%%%%%%%%%%%%%%%%%%%%%%%%%%%%%%%%%%%%%%%%%%%%%%%%%%%%%
\section{Conclusions} 
\label{Sec_Conc}
Sufficiently strong DM self-interactions provide a possible solution to the small-scale structure problems arising in the collisionless cold DM paradigm. In order to generate such a large DM self-scattering, one popular strategy is to introduce a light mediator to enhance the cross section nonperturbatively. Furthermore, this leads to the velocity-dependent DM self-interactions which help to evade the strong constraint at the scale of galaxy clusters. The simplest realization of this scenario is the weak-scale fermionic DM model with a MeV-scale light scalar or vector mediator, in which the DM relic density is controlled by the freeze-out of the DM annihilation into mediators. However, in the case with a light scalar mediator which can decay into visible SM particles like photons or electrons, the constraints from the DM relic abundance, DM direct detections and DM indirect searches strongly limit this scenario. In order to reconcile this conflict, inspired by Ref.~\cite{Ma:2017ucp,Ahmed:2017dbb,Duerr:2018mbd}, we have studied a model with a dominant fermionic DM candidate $\chi$ and a stable light mediator $S$, where the density of the latter state is depleted due to its efficient annihilation into a new scalar particle $h_2$. Consequently, the model has two advantages: first, the dominant DM $\chi$ relic density can be obtained by the conventional freeze-out mechanism where the dominant annihilation channel $\chi\bar{\chi} \to SS$ becomes invisible and can escape the DM indirect searches. Secondly, the model is even free from DM direct detection constraints since scattering of the main DM component $\chi$ with a nucleon appears at one-loop level and the light mediator is too light to have any observable signals. 

{We have fixed $s_\theta = 10^{-4.5}$, $m_2 = 1.5$~MeV,  $\kappa_{H\phi} = 3.6 \times 10^{-4}$ and set $\kappa_{HS} = 0$ to allow for the dark sector decoupling before QCD phase transition and to} avoid the limits on the properties of the dark Higgs $h_2$ from the CMB, supernovae and BBN observations. Note that even though the main annihilation channel for the freeze-out of the dominant DM particle $\chi$ is invisible, the model is still constrained by the CMB bounds on late-time electromagnetic energy injection during recombination from other visible processes such as $SS \to h_2 h_2$ and $\chi S \to \bar{\chi} h_2$ ($\bar{\chi} S \to \chi h_2$) followed by $h_2$ decays. Imposing all experimental constraints and scanning relevant regions of the parameter space, we have found (see Figs.~\ref{mp} and \ref{kap}) that there remains a large parameter region which can accommodate DM self-interactions at the level sufficient to explain the structure problems at dwarf galaxies scale while agreeing with bounds from galaxy clusters. In particular, it is found that the fraction of $S$ in the DM relic abundance is constrained to be smaller than 0.1 in this model. Also, in order to explain the cosmological small-scale structure problems, the fermionic DM mass should be lighter than ${\cal O}({\rm TeV})$, while a relatively large coupling $\kappa_{S\phi} > 10^{-4}$ is favored.

%%%%%%%%%%%%%%%%%%%%%%%%%%%%%%%%%%%%%%%%%%%%%%%%%%%%%%%%%%%%%%%%%%%%%%%%%%%
\section*{Acknowledgments}
DH would like to thank Prof. Chao-Qiang Geng for the useful discussion of symmetries in the present model. This work is supported by the National Science Centre (Poland), research project no. 2017/25/B/ST2/00191 and 2017/25/N/ST2/01312. DH is also supported by Fundação para a Ciência e a Tecnologia (FCT), within projects \textit{From Higgs Phenomenology to the Unification of Fundamental Interactions} - PTDC/FIS-PAR/31000/2017 and UID/MAT/04106/2019 (CIDMA) .

%\section*{References}
\bibliography{fsdm}
\bibliographystyle{jhep}

\end{document}